\pdfoutput=1
\documentclass[pdflatex,sn-nature,iicol]{sn-jnl}

\usepackage{graphicx}
\usepackage{multirow}
\usepackage{amsmath,amssymb,amsfonts}
\usepackage{amsthm}
\usepackage{xcolor}
\usepackage{booktabs}
\usepackage{array}
\usepackage{algorithm}
\usepackage{algorithmicx}
\usepackage{algpseudocode}
\usepackage{braket}
\usepackage{siunitx}
\usepackage{enumitem}
\usepackage{tikz}
\usetikzlibrary{positioning,arrows.meta,calc,decorations.pathmorphing,shapes.geometric,fit}

\newcommand{\eps}{\varepsilon}
\newcommand{\Gqp}{\Gamma_{\mathrm{qp}}}
\newcommand{\cfib}{c_{\mathrm{f}}}
\newcommand{\tpre}{t_{\mathrm{pre}}}
\newcommand{\tpost}{t_{\mathrm{post}}}
\newcommand{\etaint}{\eta_{\mathrm{int}}}
\newcommand{\Gint}{\bar\Gamma^{\mathrm{int}}}

\theoremstyle{thmstyleone}
\newtheorem{theorem}{Theorem}
\newtheorem{proposition}{Proposition}
\newtheorem{lemma}{Lemma}
\newtheorem{corollary}{Corollary}
\theoremstyle{thmstyletwo}
\newtheorem{remark}{Remark}
\newtheorem{hypothesis}{Hypothesis}

\setlist[itemize]{leftmargin=1.2em}
\setlist[enumerate]{leftmargin=1.5em}
\AtBeginDocument{\interdisplaylinepenalty=10000}

\AtBeginDocument{\apptocmd{\thebibliography}{\raggedright\hyphenpenalty=10000\relax}{}{}}

\begin{document}

\title[DI-QKD with Majorana Memories]{Device-Independent Quantum Key Distribution with Majorana Memories}

\author*[1]{\fnm{Noureldin} \sur{Mohamed}}\email{nomo89098@hbku.edu.qa}
\author[1]{\fnm{Saif} \sur{Al-Kuwari}}

\affil*[1]{\orgdiv{Qatar Center for Quantum Computing, College of Science and Engineering}, \orgname{Hamad Bin Khalifa University}, \orgaddress{\city{Doha}, \country{Qatar}}}

\abstract{Device-independent quantum key distribution (DI-QKD) certifies secrecy from Bell nonlocality and has been demonstrated with ions, atoms and photons. Majorana parity qubits are proposed as network memories, but their operations and noise have not been connected to device-independent security. We show that Clifford-only Majorana operations cannot violate the CHSH inequality on a heralded pair, so a node needs a non-Clifford resource, which we confine to one party's test rounds. Because quasiparticle poisoning flips the total island parity, a parity flag read before the inputs are drawn preserves device independence and turns the poisoning distance cutoff into a rate penalty; flag errors restore part of the loss in proportion to the miss probability. We derive closed-form Bell statistics for realistic heralding and anisotropic Majorana noise, instantiate a block-based finite-size key length, and find that 2025 tetron prototypes cannot yet certify a key. We identify the hardware milestones that would.}

\keywords{Device-independent quantum key distribution, Majorana zero modes, quasiparticle poisoning, Bell nonlocality, entropy accumulation, quantum networks}

\maketitle

\section{Introduction}
\label{sec:introduction}

Device-independent quantum key distribution (DI-QKD) derives the secrecy of a key from the violation of a Bell inequality, usually the Clauser-Horne-Shimony-Holt (CHSH) inequality \cite{clauser1969proposed}, and requires no model of the internal workings of the measurement devices \cite{acin2007device,pironio2009device}. Security against general coherent attacks follows from the entropy accumulation theorem (EAT) \cite{dupuis2020entropy,arnon2018practical} and its generalisations \cite{metger2024generalised}. The experimental requirements are severe. The Bell test must be free of the detection loophole, the observed CHSH value $S$ must be high, the quantum bit error rate (QBER) $Q$ must be low, and between $10^{6}$ and $10^{10}$ rounds are needed before the finite-size penalty becomes affordable \cite{arnon2018practical,murta2019towards,zapatero2023advances,primaatmaja2023security}.

Experiments have progressed quickly. In 2022, trapped ions produced a finite-size DI key across a \SI{2}{m} link \cite{nadlinger2022experimental}, single rubidium atoms separated by \SI{400}{m} reached an asymptotic key rate of 0.07 bit per entanglement event \cite{zhang2022device}, and a photonic setup reported positive asymptotic rates over up to \SI{220}{m} of fibre at a detection efficiency of \SI{87.5}{\percent} \cite{liu2022toward}. Heralded single-atom entanglement combined with single-photon interference and quantum frequency conversion has since produced finite-size DI keys over \SI{11}{km} of fibre and positive asymptotic rates at \SI{100}{km} \cite{lu2026device}. All of these matter-based systems share one architecture: long-lived memories whose entanglement is heralded by photons detected at a central station. Heralding is what makes the detection loophole a property of the local memory readout rather than of the lossy channel \cite{hensen2015loophole}. The same midpoint geometry underlies measurement-device-independent QKD \cite{lo2012measurement} and its memory-assisted variants \cite{abruzzo2014measurement,bhaskar2020experimental}, which remove trust in the measurement station but still require characterised sources; DI-QKD removes that remaining assumption at the price of a Bell violation, and it is the setting considered here.

Majorana zero modes (MZMs) in semiconductor-superconductor heterostructures \cite{lutchyn2010majorana,oreg2010helical} store quantum information in the fermion parity of spatially separated modes, which suppresses the effect of local perturbations \cite{kitaev2003fault,nayak2008non}. Tetron and hexon designs implement Pauli operators as parity measurements \cite{plugge2017majorana,karzig2017scalable}, and Clifford operations follow from measurement-only protocols \cite{bonderson2008measurement}. Interferometric single-shot parity measurement with an assignment error near \SI{1}{\percent} and a dwell time beyond \SI{1}{ms} has been reported in InAs-Al devices \cite{microsoft2025interferometric}, a tetron device has shown $Z$- and $X$-loop measurements with lifetimes $\tau_Z=\SI{12.4}{ms}$ and $\tau_X=\SI{14.5}{\micro s}$ and assignment errors of \SI{0.5}{\percent} and \SI{16}{\percent} \cite{aghaee2025distinct}, and single-shot parity readout has been achieved in a minimal Kitaev chain of quantum dots with parity lifetimes above \SI{1}{ms} \cite{vanloo2026single}. Whether these devices host topological modes remains contested \cite{aghaee2023inas,legg2026robustness}. A cryptographic analysis should therefore rest on measured parity-level quantities, which keep their meaning whatever the microscopic origin of the modes.

\subsection{Problem Statement and Hypotheses}
The question we address is whether a Majorana parity memory can serve as a DI-QKD node, and if so which hardware parameters decide the answer. Informal arguments in the literature assume that braiding can rotate the qubits into the CHSH measurement bases, that quasiparticle poisoning simply lowers the visibility, and that an audit of detection efficiencies closes the detection loophole. Each assumption fails under analysis. We therefore state five hypotheses, each grounded in established theory and each falsifiable by the analysis or by an experiment.

\begin{hypothesis}[Clifford insufficiency]
\label{hyp:clifford}
A DI-QKD node built only from topologically protected operations, meaning braiding, Pauli parity measurement and classical processing, cannot certify CHSH nonlocality from heralded entangled pairs. Falsified by exhibiting a protocol with $S>2$ that uses those operations alone on a stabilizer resource.
\end{hypothesis}

\begin{hypothesis}[Flagged poisoning]
\label{hyp:flag}
Because a poisoning event changes the total fermion parity of an island, a parity measurement performed after the herald and before the inputs are drawn removes the corresponding errors without affecting device-independent security, so poisoning limits the rate rather than the distance. Falsified by showing that the post-selection invalidates the security proof, or that the flag cannot be measured without disturbing the logical state.
\end{hypothesis}

\begin{hypothesis}[Noise anisotropy]
\label{hyp:aniso}
Majorana memory noise is anisotropic, with transverse information decaying much faster than longitudinal information, and the microscopic mechanism determines the functional form of the decay. Settings adapted to the anisotropy recover key rate that symmetric settings lose, and a Markovian rate extracted from long-time telegraph noise overestimates the damage when the underlying process is a quasi-static splitting. Falsified by showing that the optimum settings are the symmetric ones, or that the two noise models are indistinguishable at the relevant storage times.
\end{hypothesis}

\begin{hypothesis}[Finite-size dominance]
\label{hyp:finite}
For a heralded architecture whose attempt rate is bounded by the fibre round-trip time, the time to a first key is set by the number of rounds required by the finite-size analysis rather than by the asymptotic key rate. Falsified by an operating point where the asymptotic rate, not the round count, sets the time to a key.
\end{hypothesis}

\begin{hypothesis}[Interface limitation]
\label{hyp:interface}
Under parameters inferred from current devices, the limiting quantities are the $X$-basis parity lifetime and the parity-to-photon interface efficiency, not the poisoning rate that is usually quoted as the platform's bottleneck. Falsified by a parameter set in which poisoning alone prevents a key while readout and interface are adequate.
\end{hypothesis}

\subsection{Contributions}
\begin{enumerate}
\item \textbf{A no-go statement and its circumvention.} That Pauli measurements on a two-qubit stabilizer state cannot violate CHSH is elementary, and nonlocality of stabilizer states with Pauli measurements is known to require more than one pair \cite{mermin1990simple,guhne2005bell}. Proposition~\ref{prop:nogo} states the result in the form needed for Majorana hardware, including adaptive, ancilla-assisted Clifford measurements with a single shared pair, and thereby proves Hypothesis~\ref{hyp:clifford}. We then give a protocol in which only one party's test rounds consume a non-Clifford resource, and we analyse the Mermin-Peres magic square \cite{mermin1990simple,peres1990incompatible,zhen2023device} as a Pauli-native alternative whose cost is memory rather than magic.
\item \textbf{A hardware-faithful noise model.} Lemma~\ref{lem:poison} derives the logical action of a poisoning event from the Majorana algebra of a tetron. We combine it with anisotropic intrinsic switching, quasi-static level splitting, basis-dependent assignment errors and explicit two-photon \cite{barrett2005efficient} and single-photon \cite{cabrillo1999creation,humphreys2018deterministic} heralding models that include noise clicks.
\item \textbf{Just-in-time parity flagging.} Lemmas~\ref{lem:flagdamp} and \ref{lem:flag} establish Hypothesis~\ref{hyp:flag}. Conditioning on an unchanged total parity removes the damage that is first order in $\Gqp t$, so the cutoff $L_{\mathrm{qp}}\approx0.077\,\cfib/\Gqp$ becomes a rate penalty together with a second-order visibility loss whose size depends on how the poisoning couples to the four Majorana modes. For a flag that misses parity changes with probability $\eps_m$, the first-order loss returns in proportion to $\eps_m$, which sets a quantitative requirement on the flag readout.
\item \textbf{Closed-form Bell statistics and error budget.} Proposition~\ref{prop:closed} gives $S$ and $Q$ in closed form with noise-adapted settings (Proposition~\ref{prop:adapted}), validated against exact density-matrix evaluation, together with first-order sensitivities and analytic distance limits for each noise channel.
\item \textbf{Block-based finite-size security.} The protocol is stated in the bounded-block form of the entropy-accumulation analysis on which its key length rests, and its parameter estimation is defined block by block.
\item \textbf{Finite-size scaling law.} Proposition~\ref{prop:scaling} shows that the optimal test fraction and the finite-size penalty of the EAT bound both scale as $n^{-1/3}$, which we confirm numerically. Together with the computed round requirements this supports Hypothesis~\ref{hyp:finite}.
\item \textbf{Quantitative feasibility.} We compute key rates, minimum round numbers, feasibility maps and secret bits per day, and compare them with the published DI-QKD experiments. Parameters inferred from 2025 tetron prototypes give $S<2$, which supports Hypothesis~\ref{hyp:interface} and identifies the experimental milestones.
\end{enumerate}

The paper is organised as follows. Section~\ref{sec:preliminaries} reviews DI-QKD, entropy accumulation and Majorana parity qubits. Section~\ref{sec:system} defines the system and noise model. Section~\ref{sec:nonlocality} treats nonlocality with Majorana primitives, Section~\ref{sec:protocol} the protocol, and Section~\ref{sec:security} the security analysis. Section~\ref{sec:closedform} derives the closed-form statistics and the error budget, Section~\ref{sec:results} presents the numerical results, and Section~\ref{sec:discussion} discusses limitations and outlook.

\section{Preliminaries}
\label{sec:preliminaries}

\subsection{CHSH-Based DI-QKD}
Alice and Bob hold uncharacterised devices. In each round Alice selects $x\in\{0,1\}$ and Bob selects $y\in\{0,1,2\}$, and the devices return $a,b\in\{\pm1\}$. Test rounds use $x,y\in\{0,1\}$ and estimate
\begin{equation}
\begin{gathered}S=E_{00}+E_{01}+E_{10}-E_{11},\\ E_{xy}=\sum_{a,b}ab\,P(a,b|x,y),\end{gathered}
\label{eq:chsh}
\end{equation}
which is equivalent to the winning probability $\omega=\tfrac12+S/8$ of the CHSH game. Local models obey $S\le2$ and quantum theory allows $S\le2\sqrt2$ \cite{cirelson1980quantum}. Key rounds use $(x,y)=(0,2)$ and define $Q=P(a\neq b|0,2)$. Against collective attacks in the asymptotic limit, the key rate per round satisfies \cite{pironio2009device}
\begin{equation}
r_\infty(S,Q)=1-h\!\left(\tfrac12+\tfrac12\sqrt{(S/2)^2-1}\right)-h(Q),
\label{eq:pironio}
\end{equation}
with $h$ the binary entropy. The same entropy bound serves as the min-tradeoff function of the EAT and therefore also covers coherent attacks \cite{arnon2018practical}. For Werner-type noise, where $S=2\sqrt2(1-2Q)$, Eq.~\eqref{eq:pironio} is positive for $Q<\SI{7.15}{\percent}$. Noisy preprocessing raises this threshold \cite{ho2020noisy}; our implementation of that bound gives \SI{8.08}{\percent}. Asymmetric CHSH inequalities \cite{schwonnek2021device}, numerical entropy bounds based on semidefinite relaxations \cite{brown2024device,tan2022improved} and advantage distillation \cite{tan2020advantage} improve the tolerance further. We work with Eq.~\eqref{eq:pironio} because it is analytic, tight among bounds that depend on $S$ alone, and conservative with respect to these refinements, which use more of the observed statistics. Where a refinement changes a number by more than a few per cent we quote it explicitly.

The detection loophole enters in two ways. Data rejection can reproduce nonlocal statistics when detection is inefficient \cite{pearle1970hidden}. In a direct photonic test with maximally entangled states and no-click events assigned a fixed outcome, a CHSH violation requires a detection efficiency above $2/(1+\sqrt2)\approx\SI{82.8}{\percent}$ \cite{garg1987detector}; partially entangled states lower the threshold \cite{eberhard1993background}, which is how the photonic loophole-free tests were realised \cite{giustina2015significant,shalm2015strong}. In an event-ready configuration the herald precedes the choice of inputs, so channel loss reduces the rate but does not open the loophole, and the efficiency that matters is that of the local measurement \cite{hensen2015loophole,zhang2022device,lu2026device}.

\subsection{Entropy Accumulation}
The EAT bounds the smooth min-entropy of the outputs of $n$ sequential rounds \cite{dupuis2020entropy}. Let each round be described by a channel $\mathcal{M}_i$ that produces an output $A_i$, side information accessible to Eve, and a statistics register $C_i$, and let $f$ be an affine min-tradeoff function, meaning that $f(p)$ lower-bounds the conditional von Neumann entropy of a single round whose statistics are $p$. If the Markov conditions of \cite{dupuis2020entropy} hold, then conditioned on an event $\Omega$ in which the observed frequency of $C$ lies in a set with $f\ge t$,
\begin{equation}
H_{\min}^{\eps_s}(A_1^n|E)_{\rho|\Omega}\;>\;n\,t-v\sqrt n ,
\label{eq:eat}
\end{equation}
where
\begin{multline}
v=2\left(\log_2(1+2d_A)+\lceil\lVert\nabla f\rVert_\infty\rceil\right)\\
\times\sqrt{1-2\log_2(\eps_s\,\rho[\Omega])}
\label{eq:eatv}
\end{multline}
and $d_A$ is the dimension of the output alphabet. Section~\ref{sec:security} instantiates Eq.~\eqref{eq:eat} for our protocol. Rényi versions of the theorem \cite{metger2024generalised}, used in the \SI{100}{km} experiment of \cite{lu2026device}, reduce the second-order term by a large factor, so the round requirements reported below are upper bounds.

\subsection{Majorana Parity Qubits}
A tetron hosts four MZMs with Hermitian operators $\gamma_j$ obeying $\{\gamma_j,\gamma_k\}=2\delta_{jk}$. At fixed total parity $P=-\gamma_1\gamma_2\gamma_3\gamma_4$ the logical operators are the bilinears
\begin{equation}
Z_L=i\gamma_1\gamma_2,\qquad X_L=i\gamma_2\gamma_3,\qquad Y_L=i\gamma_1\gamma_3 ,
\label{eq:logical}
\end{equation}
which satisfy the Pauli algebra and commute with $P$ \cite{plugge2017majorana,karzig2017scalable}. Pauli measurements are realised as interferometric parity measurements through coupled quantum dots, read out dispersively \cite{microsoft2025interferometric,vanloo2026single}, and the Clifford group is generated by braids or by sequences of parity measurements \cite{nayak2008non,bonderson2008measurement}. Operations outside the Clifford group require either an unprotected geometric phase gate \cite{karzig2016universal} or the injection of a distilled magic state \cite{bravyi2005universal}.
\begin{table}[!tb]
\centering
\caption{Reported parity-level figures of merit used to anchor the parameter tiers. Their topological interpretation is debated \cite{legg2026robustness}; the model below uses them as phenomenological inputs.}
\label{tab:sota}
\begin{tabular}{@{}>{\raggedright\arraybackslash}p{3.7cm}>{\raggedright\arraybackslash}p{2.5cm}l@{}}
\toprule
Quantity & Reported value & Ref.\\
\midrule
Single-shot parity assignment error & $\approx\SI{1}{\percent}$ & \cite{microsoft2025interferometric}\\
Parity dwell time, single wire & $>\SI{1}{ms}$ & \cite{microsoft2025interferometric}\\
Parity lifetime, Kitaev chain & $>\SI{1}{ms}$ & \cite{vanloo2026single}\\
Tetron $Z$-loop lifetime $\tau_Z$ & \SI{12.4}{ms}, poisoning & \cite{aghaee2025distinct}\\
Tetron $X$-loop lifetime $\tau_X$ & \SI{14.5}{\micro s}, splitting & \cite{aghaee2025distinct}\\
Assignment errors $\eps_Z$, $\eps_X$ & \SI{0.5}{\percent}, \SI{16}{\percent} & \cite{aghaee2025distinct}\\
Integration time for ${\rm SNR}=1$ & \SI{3.7}{\micro s} & \cite{microsoft2025interferometric}\\
\bottomrule
\end{tabular}
\end{table}
\begin{figure*}[!t]
\centering
\begin{tikzpicture}[font=\scriptsize, >={Stealth[length=4pt]},
  node box/.style={draw, rounded corners=2pt, minimum width=3.3cm, minimum height=2.6cm, fill=blue!3},
  comp/.style={draw, rounded corners=1pt, fill=white, minimum width=2.95cm, minimum height=0.36cm, inner sep=1pt}]
\node[node box] (A) at (0,0) {};
\node[anchor=north] at (A.north) {\textbf{Alice}};
\node[comp] (Atet) at (0,0.55) {tetron memory};
\node[comp] (Aint) at (0,0.05) {parity$\to$photon, $\etaint$};
\node[comp] (Arng) at (0,-0.45) {RNG $x$, flag $F_A$};
\node[comp] (Acl) at (0,-0.95) {classical processing};
\node[node box] (B) at (8.8,0) {};
\node[anchor=north] at (B.north) {\textbf{Bob}};
\node[comp] (Btet) at (8.8,0.55) {tetron memory};
\node[comp] (Bint) at (8.8,0.05) {parity$\to$photon, $\etaint$};
\node[comp] (Brng) at (8.8,-0.45) {RNG $y$, flag $F_B$};
\node[comp] (Bcl) at (8.8,-0.95) {classical processing};
\node[draw, fill=orange!8, rounded corners=2pt, minimum width=1.6cm, minimum height=1.0cm, align=center] (H) at (4.4,0.05) {heralding\\station\\(untrusted)};
\draw[thick, blue!60!black, decorate, decoration={snake, amplitude=0.6pt, segment length=4pt}] (A.east |- Aint.east) -- node[above=1pt]{fibre $L/2$} (H.west);
\draw[thick, blue!60!black, decorate, decoration={snake, amplitude=0.6pt, segment length=4pt}] (B.west |- Bint.west) -- node[above=1pt]{fibre $L/2$} (H.east);
\draw[->, dashed] (H.south) |- node[pos=0.75, below=1pt]{herald $h$} (A.east |- Acl.east);
\draw[->, dashed] (H.south) |- (B.west |- Bcl.west);
\draw[<->, gray] (A.south) -- ++(0,-0.3) -- node[below]{authenticated classical channel} ($(B.south)+(0,-0.3)$) -- (B.south);
\end{tikzpicture}
\caption{Architecture. Tetron memories emit parity-conditioned photons towards an untrusted midpoint station whose click pattern heralds entanglement. After the herald arrives, each node measures the total island parity and only then draws its measurement input from a trusted RNG. The RNGs and the classical processing are trusted; the memories, interfaces, fibre and station are not.}
\label{fig:architecture}
\end{figure*}

Three noise processes are relevant. Quasiparticle poisoning exchanges a fermion with the environment and therefore changes the island parity \cite{rainis2012majorana,karzig2021quasiparticle}; in a floating island the charging energy suppresses it by making the addition of a fermion energetically costly \cite{plugge2017majorana,karzig2017scalable}, and the residual splitting between Majorana modes falls exponentially with wire length \cite{albrecht2016exponential}. Residual hybridisation of MZMs at finite wire length produces an energy splitting that dephases superpositions of the corresponding parity \cite{knapp2018dephasing,albrecht2016exponential}. Measurement introduces assignment errors that depend on the loop being interrogated \cite{aghaee2025distinct}. Table~\ref{tab:sota} lists the reported values that anchor our parameter tiers, and the published device roadmap for these architectures \cite{aasen2025roadmap} gives the context for the target values used later.

\section{System Model}
\label{sec:system}

\subsection{Architecture and Trust Assumptions}
Fig.~\ref{fig:architecture} shows the architecture. Each node holds one or more tetron memories in a dilution refrigerator, a parity-to-photon interface, a trusted random-number generator (RNG) and trusted classical processing. The nodes are joined by a fibre of total length $L$ through a central heralding station, and an authenticated classical channel connects the parties. We adopt the standard DI trust model \cite{acin2007device,arnon2018practical,portmann2022security}:
\begin{itemize}
\item quantum theory is valid and the laboratories are isolated, so that no information leaves a laboratory except through the classical channel;
\item the RNGs are uniform and independent of the devices, and the classical post-processing is trusted;
\item the devices operate sequentially, and the input of a round is chosen after all registers of previous rounds have been fixed;
\item the memories, the interfaces, the fibre and the heralding station are uncharacterised and may have been prepared by the adversary, who may keep quantum side information and exploit device memory.
\end{itemize}
Spacelike separation of the measurement events is not needed for DI-QKD security; it matters for a foundational loophole-free Bell test, where the no-signalling condition must be enforced by relativistic causality rather than by shielding \cite{zapatero2023advances}. We also do not assume that the devices are honest: every statement below holds for arbitrary device behaviour consistent with the observed statistics.

\subsection{Parity-to-Photon Interface}
\label{sec:interface}
No optical interface to a Majorana parity qubit has been demonstrated. Two routes have been proposed. The parity can be coupled to a microwave cavity, either dispersively through a parity-dependent quantum capacitance or by parametric modulation of the light-matter coupling \cite{contamin2021hybrid}, or it can be transferred to a quantum-dot spin qubit \cite{leijnse2011quantum}. Either route must then be followed by microwave-to-optical transduction, which remains far from unit efficiency and adds noise photons \cite{lauk2020perspectives,mirhosseini2020superconducting}. We therefore treat the interface as a black box with three parameters: the efficiency $\etaint$ from memory to a telecom photon in the fibre, the visibility $V$ of the heralded coherence, which absorbs photon distinguishability and optical phase noise, and the probability $p_n$ of a noise click per detection window, which absorbs detector dark counts and added transducer noise. The transmission from a node to the central detector is
\begin{equation}
\eta=\etaint\,\eta_{\mathrm{det}}\,10^{-\alpha L/20},
\label{eq:eta}
\end{equation}
with fibre attenuation $\alpha=\SI{0.2}{dB/km}$ and detector efficiency $\eta_{\mathrm{det}}$. Treating the interface in this way keeps the analysis independent of which physical route is eventually realised, and it makes $\etaint$ an explicit requirement rather than a hidden assumption.

\subsection{Heralding Schemes}
\label{sec:heralding}
After the Pauli-frame correction implied by the click pattern, which is a relabelling of classical outcomes and needs no physical gate, the heralded two-qubit state has a diagonal correlation matrix $\mathrm{diag}(t_{xx},t_{yy},t_{zz})$. The two schemes give the following expressions, which are derived at the end of this subsection.

\emph{Two-photon (Barrett-Kok) heralding.} Two emission rounds separated by a logical flip succeed with probability $P_{\mathrm{T}}=\eta^2/2$ \cite{barrett2005efficient}. A false herald, in which a genuine click in one round coincides with a noise click in the other, occurs with probability $P_{\mathrm{F}}\simeq\eta p_n$, to leading order for $p_n\ll\eta$, and leaves a $Z$-anticorrelated product state. With $\zeta=P_{\mathrm F}/(P_{\mathrm T}+P_{\mathrm F})$,
\begin{equation}
\begin{gathered}P_h=P_{\mathrm T}+P_{\mathrm F},\quad t_{zz}=1-2\zeta,\\ t_{xx}=-t_{yy}=(1-\zeta)V.\end{gathered}
\label{eq:bk}
\end{equation}

\emph{Single-photon interference.} Each node prepares $\sqrt{1-\alpha_b}\ket{0}+\sqrt{\alpha_b}\ket{1}$, where $\ket{1}$ emits a photon \cite{cabrillo1999creation,humphreys2018deterministic}. For threshold detectors the unnormalised weights of the heralded components are $w_\Psi=\alpha_b(1-\alpha_b)\eta$ for the entangled component, $w_{11}=\alpha_b^2(\eta-\eta^2/2)$ for double emission with one photon lost, and $w_{00}=(1-\alpha_b)^2p_n$ for a noise click. With $W=w_\Psi+w_{11}+w_{00}$,
\begin{multline}
P_h=2W,\qquad t_{zz}=\frac{w_\Psi-w_{11}-w_{00}}{W},\\ t_{xx}=-t_{yy}=\frac{Vw_\Psi}{W}.
\label{eq:spi}
\end{multline}
The success probability of this scheme is first order in $\eta$, which gives the square-root scaling with total channel transmission familiar from single-photon and twin-field protocols \cite{kolodynski2020device,lucamarini2018overcoming}, at the price of an error floor set by $\alpha_b$. It also requires the optical phase difference of the two arms to be stable over the fibre length, which two-photon heralding does not; we account for this with an additional factor $V_\phi\le1$ multiplying $V$ in Eq.~\eqref{eq:spi}, set $V_\phi=1$ in the tiers, and quantify the effect of $V_\phi<1$ in Sec.~\ref{sec:results}. We optimise $\alpha_b$ at each distance.

\begin{proof}[Derivation of Eqs.~\eqref{eq:bk} and \eqref{eq:spi}]
Both schemes are analysed with threshold detectors at the central station, a beam splitter that erases the which-path information, and independent loss $\eta$ in each arm as given by Eq.~\eqref{eq:eta}.

\emph{Single-photon interference.} Each node prepares $\sqrt{1-\alpha_b}\ket{0}+\sqrt{\alpha_b}\ket{1}$, and $\ket{1}$ emits one photon. In the one-photon subspace the amplitude reaching the station is $\sqrt{\alpha_b(1-\alpha_b)\eta}\,(\ket{10}\ket{1_a}+\ket{01}\ket{1_b})$, and the beam splitter maps $\ket{1_a}\to(\ket{c}+\ket{d})/\sqrt2$ and $\ket{1_b}\to(\ket{c}-\ket{d})/\sqrt2$. Projecting on a click in $c$ alone leaves the entangled component with weight $\alpha_b(1-\alpha_b)\eta$. If both nodes are bright, which happens with probability $\alpha_b^2$, a click in $c$ alone occurs when both photons arrive and bunch, with probability $\eta^2/2$, or when exactly one photon survives and reaches $c$, with probability $\eta(1-\eta)$; the total is $\eta-\eta^2/2$ and the post-measurement state is $\ket{11}$. If both nodes are dark, a noise click contributes weight $(1-\alpha_b)^2p_n$ with the state $\ket{00}$. Normalising and adding the same contribution from detector $d$ gives Eq.~\eqref{eq:spi}, where the visibility $V$ multiplies the coherence of the entangled component to account for photon distinguishability and phase noise.

\emph{Barrett-Kok.} Both nodes start in an equal superposition and emit; a single click is recorded, a logical flip is applied, and the nodes emit again. Acceptance requires one click in each round. Two genuine photons produce this pattern with probability $\eta^2/2$, the factor one half arising from the click patterns that are discarded. A false herald requires one genuine click and one noise click, which contributes $P_{\mathrm F}\simeq\eta p_n$ with the two orderings included, and leaves the two memories in a product state that is anticorrelated in $Z_L$ after the Pauli-frame correction. Equation~\eqref{eq:bk} follows.
\end{proof}

\subsection{Memory Noise}
\label{sec:memory}

\begin{lemma}[Action of poisoning on a tetron]
\label{lem:poison}
Model a quasiparticle-tunnelling event that couples to mode $j$ by $\rho\mapsto\gamma_j\rho\gamma_j$. On the logical operators of Eq.~\eqref{eq:logical}, the operators $\gamma_1,\gamma_2,\gamma_3,\gamma_4$ act as $X_L$, $Y_L$, $Z_L$ and the identity, and each of them anticommutes with the total parity $P$. If mode $j$ is affected with probability $w_j$, one event multiplies the Bloch components by $1-2(w_1+w_2)$ for $Z_L$, by $1-2(w_2+w_3)$ for $X_L$ and by $1-2(w_1+w_3)$ for $Y_L$. For a Poisson process of rate $\Gqp$ with $w_j=1/4$, every component decays as $e^{-\Gqp t}$, which is logical depolarisation, and the probability that the total parity has changed is $\tfrac12(1-e^{-2\Gqp t})$.
\end{lemma}
\begin{proof}
A single Majorana operator anticommutes with every bilinear that contains it and commutes with every bilinear that does not, because it anticommutes with each of the other three operators. Hence $\gamma_1$ anticommutes with $Z_L$ and $Y_L$ and commutes with $X_L$, which is the conjugation action of $X_L$; the other three cases follow in the same way. Since $\gamma_j$ anticommutes with three of the four factors of $P$, it anticommutes with $P$. If a component is flipped by events of total rate $\lambda$, the probability of an odd number of such events in time $t$ is $\tfrac12(1-e^{-2\lambda t})$, so the component decays as $e^{-2\lambda t}$. For $Z_L$ we have $\lambda=(w_1+w_2)\Gqp$, and $\lambda=\Gqp/2$ when $w_j=1/4$.
\end{proof}

Two physical points about the logical space deserve mention. The logical states of Eq.~\eqref{eq:logical} share one total parity, so the superpositions required by both heralding schemes are allowed by fermion-parity superselection. After an unflagged poisoning event the island sits in the opposite total-parity sector, where the bilinears of Eq.~\eqref{eq:logical} are still defined and still return $\pm1$; this is what the Pauli description of the lemma assumes, and a device whose readout fails in that sector is covered by the binning of Corollary~\ref{cor:eff}.

Lemma~\ref{lem:poison} has two consequences. First, describing poisoning as a logical bit flip with probability $\tfrac12(1-e^{-\Gqp t})$ is incorrect on both counts: the logical action is a random Pauli, and the exponent carries a factor of two. Second, every poisoning event leaves a classical trace in $P$, which Sec.~\ref{sec:protocol} exploits.

Residual hybridisation and thermal excitation of low-energy modes change individual pair parities without changing $P$ \cite{knapp2018dephasing,aghaee2025distinct}. So do quasiparticles that are already present in the island and scatter off the Majorana modes, since they act through bilinears $i\gamma_a\gamma_b$ rather than single operators \cite{karzig2021quasiparticle}. None of these processes is visible to a total-parity measurement. We treat them together in two limits.

\emph{Markovian limit.} Switching events form Poisson processes with rates $\Gamma^{\mathrm{int}}_x$, which flips $Z_L$, and $\Gamma^{\mathrm{int}}_z$, which flips $X_L$. This is the appropriate description when the telegraph noise is fast compared with the storage time.

\emph{Quasi-static limit.} A residual splitting $\delta$ between the two parity states of a loop generates the unitary $e^{-i\delta t\,Z_L/2}$. If $\delta$ is constant within a round but drawn from a zero-mean Gaussian of standard deviation $\sigma_\delta$ across rounds, which is the standard description of slow charge or flux noise, then averaging over $\delta$ multiplies the transverse components by $e^{-\sigma_\delta^2t^2/2}$ per node and leaves the longitudinal component untouched. A static and known splitting is not a decoherence channel at all, since it can be absorbed into the definition of the measurement axis.

The effect of poisoning on the observed correlations depends on whether the total parity is checked, and Lemma~\ref{lem:flagdamp} gives the two cases in closed form.

\begin{lemma}[Damping with and without parity conditioning]
\label{lem:flagdamp}
Let poisoning events form a Poisson process of rate $\Gqp$ over a window of length $t$, each flipping a given Bloch component with probability $u$ as in Lemma~\ref{lem:poison}. Without conditioning the component is multiplied by $e^{-2u\Gqp t}$. Conditioned on the total parity being unchanged at the end of the window, an event of probability $P_{\mathrm{even}}=\tfrac12(1+e^{-2\Gqp t})$, it is multiplied by
\begin{equation}
\Phi(u,t)=\frac{\cosh\!\big[(1-2u)\Gqp t\big]}{\cosh(\Gqp t)} .
\label{eq:phi}
\end{equation}
For uniform weights, $u=1/2$, this equals $1/\cosh(\Gqp t)=1-\tfrac12(\Gqp t)^2+O(t^4)$, and for a single dominant mode, $u\in\{0,1\}$, it equals one.
\end{lemma}
\begin{proof}
Let $N$ be the number of events, which is Poisson distributed with mean $\Gqp t$. Each event multiplies the component by $-1$ with probability $u$, so the unconditional damping is $\mathbb{E}[(1-2u)^N]=e^{-\Gqp t[1-(1-2u)]}=e^{-2u\Gqp t}$. Every event flips the total parity by Lemma~\ref{lem:poison}, so an unchanged parity is equivalent to $N$ being even, and $\sum_{N\text{ even}}e^{-\Gqp t}(\Gqp t)^N a^N/N!=e^{-\Gqp t}\cosh(a\Gqp t)$. Dividing by the same sum at $a=1$ gives Eq.~\eqref{eq:phi} and $P_{\mathrm{even}}=e^{-\Gqp t}\cosh(\Gqp t)$.
\end{proof}

The flag therefore always removes the damage that is first order in $\Gqp t$, and the residual term is quadratic. Whether even that term is present depends on the microscopic weights: two events that both act on the same Majorana mode return the qubit to its initial state, so a poisoning process dominated by one mode is removed completely, while uniform weights leave $1-\tfrac12(\Gqp\tpre)^2$. Sec.~\ref{sec:res_flag} evaluates both cases.

\emph{Imperfect flags.} Lemma~\ref{lem:flagdamp} assumes perfect parity discrimination. Suppose instead that the flag misses a changed total parity with probability $\eps_m$ and raises a false alarm on an unchanged parity with probability $\eps_f$, independently in each round. Repeating the argument of Lemma~\ref{lem:flagdamp} with the odd-event branch kept with weight $\eps_m$ and the even-event branch with weight $1-\eps_f$ gives, with $x=\Gqp t$ and $a=1-2u$, the damping
\begin{equation}
\Phi_{\eps}(u,t)=\frac{(1-\eps_f)\cosh(ax)+\eps_m\sinh(ax)}{(1-\eps_f)\cosh x+\eps_m\sinh x}
\label{eq:phiimp}
\end{equation}
and a kept fraction $e^{-x}\big[(1-\eps_f)\cosh x+\eps_m\sinh x\big]$ per node, which reduce to Eq.~\eqref{eq:phi} and to $P_{\mathrm{even}}$ for $\eps_m=\eps_f=0$. Expanding to second order,
\begin{equation}
\Phi_{\eps}(u,t)\simeq1-\frac{\eps_m}{1-\eps_f}\,2ux-2u(1-u)x^2 .
\label{eq:phiimpexp}
\end{equation}
\begin{proof}[Derivation of Eqs.~\eqref{eq:phiimp} and \eqref{eq:phiimpexp}]
With $N$ Poisson distributed with mean $x$, the unnormalised contributions of the even and odd branches to $\mathbb{E}[a^N]$ are $e^{-x}\cosh(ax)$ and $e^{-x}\sinh(ax)$. An imperfect flag keeps the even branch with probability $1-\eps_f$ and the odd branch with probability $\eps_m$, so the kept weight is $e^{-x}[(1-\eps_f)\cosh(ax)+\eps_m\sinh(ax)]$ and the kept probability is the same expression at $a=1$. Their ratio is Eq.~\eqref{eq:phiimp}. Expanding $\cosh(ax)\simeq1+a^2x^2/2$ and $\sinh(ax)\simeq ax$ and using $(1-a^2)/2=2u(1-u)$ gives Eq.~\eqref{eq:phiimpexp}.
\end{proof}
A missed flag therefore restores the first-order loss $2u\Gqp t$ of the unflagged protocol in proportion to $\eps_m/(1-\eps_f)$, whereas false alarms leave the correlations unchanged and only reduce the rate by $(1-\eps_f)^2$ per memory pair. The single-mode case, in which the perfect flag removes all damage, is the most sensitive to $\eps_m$, because the residual quadratic term that dominates for uniform weights is then absent. A flag readout that is not perfectly quantum non-demolition adds a dephasing of the logical bilinears, which enters in the same way as the unflagged rates of Eq.~\eqref{eq:memory}. Eq.~\eqref{eq:phiimp} and the kept fraction agree with a Monte-Carlo simulation of $2\times10^6$ Poisson event histories per parameter set to within $2\times10^{-3}$, the sampling error, and Sec.~\ref{sec:res_flag} quantifies the effect on the reach.

Collecting the three processes, the per-node damping factors for a qubit held for $\tpre$ before the parity flag and $\tpost$ afterwards, with $t=\tpre+\tpost$ and $u_Z=w_1+w_2$, $u_X=w_2+w_3$, are
\begin{align}
f_z&=\Phi(u_Z,\tpre)\,e^{-2u_Z\Gqp\tpost}\,e^{-2\Gamma^{\mathrm{int}}_x t},\nonumber\\
f_x&=\Phi(u_X,\tpre)\,e^{-2u_X\Gqp\tpost}\,e^{-2\Gamma^{\mathrm{int}}_z t}\,e^{-\sigma_\delta^2t^2/2},
\label{eq:memory}
\end{align}
when the flag of Sec.~\ref{sec:protocol} is used, in which case each node keeps a fraction $P_{\mathrm{even}}=\tfrac12(1+e^{-2\Gqp\tpre})$ of the heralded rounds, and $\Phi(u,\tpre)e^{-2u\Gqp\tpost}$ is replaced by $e^{-2u\Gqp t}$ otherwise; with an imperfect flag, $\Phi$ is replaced by $\Phi_{\eps}$ of Eq.~\eqref{eq:phiimp}.

\subsection{Measurement, Timing and Multiplexing}
The $Z_L$ and $X_L$ parity measurements have assignment errors $\eps_Z$ and $\eps_X$, and we write $c_k=1-2\eps_k$. Readouts that fail to converge occur with probability $1-\eta_r$ and are assigned the outcome $+1$, which Corollary~\ref{cor:eff} treats. Bob's test settings need a rotation that lies outside the Clifford group, as shown in Sec.~\ref{sec:nonlocality}; we model it as a depolarising channel of strength $\eps_T$ followed by a $Z_L$ readout, and treat a systematic angle error separately in Sec.~\ref{sec:discussion}.

The storage time before the flag is $\tpre=L/\cfib$ for single-photon heralding and $\tpre=L/\cfib+t_{\mathrm{ops}}$ for Barrett-Kok, where $\cfib=\SI{2e8}{m/s}$ is the group velocity in fibre and $t_{\mathrm{ops}}$ the time for initialisation and emission. The photon travels $L/2$ to the station and the herald returns over $L/2$, so the latency is $L/\cfib$ in both cases. A node cannot start a new attempt before the previous herald has been resolved, because a success would otherwise be destroyed, so a single memory pair performs $R_{\mathrm{att}}=1/(L/\cfib+t_{\mathrm{ops}})$ attempts per second for single-photon heralding and $1/(L/\cfib+2t_{\mathrm{ops}})$ for Barrett-Kok. With $M$ memory pairs operated in parallel on distinguishable optical modes the attempt rate is multiplied by $M$, and the analysis below treats $M$ as a design variable.

\section{Nonlocality with Majorana Primitives}
\label{sec:nonlocality}

\subsection{A Clifford No-Go Theorem for CHSH}

\begin{proposition}[Clifford no-go]
\label{prop:nogo}
Let $\rho_{AB}$ be any convex mixture of two-qubit stabilizer states, with one qubit held by each party.
\begin{enumerate}[label=(\roman*)]
\item If each party applies a local Clifford unitary, which includes any braid, measures a Pauli observable selected by the input, and post-processes the outcome by an arbitrary local classical map, which includes assignment errors and the binning of failed readouts, then the behaviour $P(a,b|x,y)$ admits a local hidden-variable model, and in particular $|S|\le2$ for all eight CHSH expressions.
\item The same holds if each party may in addition prepare local ancillas in stabilizer states, apply Clifford gates, and perform an adaptive sequence of Pauli measurements on its shared qubit and ancillas, with later choices depending on earlier outcomes and on the input.
\end{enumerate}
\end{proposition}
\begin{proof}
(i) Local Clifford unitaries map Pauli observables to Pauli observables and stabilizer states to stabilizer states, so they may be absorbed into the choice of observable. Local classical post-processing maps local behaviours to local behaviours, and locality is preserved under convex mixtures, so it suffices to prove that projective Pauli measurements $A_x,B_y\in\pm\{X,Y,Z\}$ on a pure two-qubit stabilizer state give a local behaviour. Such a state is either a product state, in which case the behaviour is local by construction, or it is equivalent up to local Cliffords to $\ket{\Phi^+}$. For $\ket{\Phi^+}$ one has $\langle\sigma\otimes\sigma'\rangle=0$ for distinct Pauli operators $\sigma\neq\sigma'$, while $\langle XX\rangle=\langle ZZ\rangle=-\langle YY\rangle=1$, and all single-party marginals vanish. If $A_0=\pm A_1$, each CHSH expression collapses to $\langle A_0\otimes(B_0\pm B_1)\rangle$, whose modulus is at most 2 because $\lVert B_0\pm B_1\rVert\le2$. Otherwise $A_0$ and $A_1$ are different Pauli operators up to sign, each $B_y$ agrees with at most one of them, and at most two of the four correlators are non-zero, each of modulus at most one, so again $|S|\le2$. Because the marginals vanish, the correlators determine the behaviour through $P(a,b|x,y)=\tfrac14(1+abE_{xy})$, and by Fine's theorem the eight CHSH inequalities are necessary and sufficient for locality in the two-input, two-output scenario.

(ii) By convexity it suffices to take a pure global state. Throughout the protocol the global state is a stabilizer state, and its entanglement across the Alice-Bob cut is an integer number of ebits that local operations cannot increase; it starts at one. A bipartite stabilizer state carrying one ebit is equivalent under local Clifford operations to $\ket{\Phi^+}$ on one qubit per side together with local stabilizer states \cite{fattal2004entanglement,audenaert2005entanglement}. In that frame each Pauli measurement of Alice reads $\sigma_a\otimes m$, with $\sigma_a$ acting on her half of $\ket{\Phi^+}$ and $m$ on her local qubits. Three cases arise. If $\sigma_a=\mathbb{1}$, the measurement acts on a local state and yields local randomness. If $\sigma_a\neq\mathbb{1}$ and $m$ commutes with the stabilizer of the local state, then $m$ acts on it as a fixed sign, the outcome equals that sign times the outcome of $\sigma_a$ on the shared pair, and the ebit is consumed. If $m$ anticommutes with an element of the local stabilizer, the outcome is uniformly random and uncorrelated with every Pauli measurement of Bob, because $\sigma_a m\otimes\sigma_b$ is not in the stabilizer group up to sign for any $\sigma_b$; the ebit then survives in a new local Clifford frame. By induction, Alice's record consists of local randomness, the input, and at most one outcome of a Pauli measurement on $\ket{\Phi^+}$, whose choice depends only on her input and local randomness, and the same holds for Bob. The two parties' operations commute, so their order is irrelevant. Conditioned on the independent local randomness of both parties, the behaviour is of the form treated in (i) and is local, and mixing over independent local randomness preserves locality.
\end{proof}

\begin{remark}
\label{rem:nogo}
The proposition is tight in three respects, and its content for a single pair is elementary; its role here is to close the loophole that measurement-based Majorana operations might evade it through ancillas and adaptivity, which part (ii) excludes. It fails for non-stabilizer states: the Pauli settings $A_0=B_0=Z$ and $A_1=B_1=X$ attain $2\sqrt2$ on a Bell state rotated by $e^{-i\pi Y/8}$ on one side, since the corresponding Bell operator squares to $4\mathbb{1}+4\,Y\otimes Y$. Preparing that state, however, consumes the same non-Clifford resource that the measurements were supposed to avoid. The proposition also fails for more than one shared pair, since the second pair can be measured after the first has been consumed: the magic-square game is won with certainty using Pauli measurements on two Bell pairs, as discussed in Sec.~\ref{sec:ms}, and stabilizer states of more qubits violate suitable Bell inequalities under Pauli measurements \cite{guhne2005bell}. Finally, efficient classical simulability of Clifford circuits \cite{aaronson2004improved} does not by itself imply locality, so a separate argument of the kind given above is needed.
\end{remark}

Proposition~\ref{prop:nogo} settles Hypothesis~\ref{hyp:clifford} and has a direct design consequence. Heralded states of the form \eqref{eq:bk} or \eqref{eq:spi} are mixtures of stabilizer states, so a node restricted to protected operations produces local statistics and certifies no key. Braiding cannot rotate a Majorana qubit into a CHSH eigenbasis, and no amount of hardware improvement changes that conclusion.

\subsection{Resource-Minimal CHSH Protocol}
We place the non-Clifford resource where it costs least. Alice measures $A_0=Z_L$, used in key and test rounds, and $A_1=X_L$, used in test rounds; both are native parity measurements. Bob measures $B_2=Z_L$ in key rounds and
\begin{equation}
B_{0,1}=\cos\theta\,Z_L\pm\sin\theta\,X_L=R_y(\mp\theta)^\dagger Z_L R_y(\mp\theta)
\label{eq:bob}
\end{equation}
in test rounds, implemented as a rotation followed by a $Z_L$ readout. At $\theta=\pi/4$ the rotation is Clifford-equivalent to a $T$ gate, which can be obtained from a geometric phase gate along a path that is protected to first order in the control errors \cite{karzig2016universal} or by injecting a distilled magic state \cite{bravyi2005universal}. Three properties make this arrangement economical. The resource is consumed only in Bob's test rounds, that is at rate $\mu$ per heralded round, where $\mu$ is the test fraction, which Sec.~\ref{sec:security} shows to be of order $10^{-2}$. The rotation error $\eps_T$ enters $S$ but not $Q$, since key rounds are Clifford. The magic-state factory can run in parallel with the network operation, because the resource is consumed at a rate far below the local gate rate.

\subsection{Settings Adapted to the Noise Anisotropy}

\begin{proposition}[Adapted CHSH settings]
\label{prop:adapted}
Consider a state with correlation matrix $\mathrm{diag}(T_x,T_y,T_z)$ and vanishing $Z$-$X$ cross-correlations, and define the effective correlations $a=c_Z^2T_z$ and $b=c_Xc_ZT_x$ that include assignment errors. With the settings of Eq.~\eqref{eq:bob},
\begin{equation}
S(\theta)=2(1-\eps_T)\,(a\cos\theta+b\sin\theta),
\label{eq:Stheta}
\end{equation}
which is maximised at $\tan\theta^\star=b/a$ and gives $S^\star=2(1-\eps_T)\sqrt{a^2+b^2}$. For ideal readout and $|T_z|,|T_x|\ge|T_y|$ this coincides with the Horodecki maximum $2\sqrt{T_z^2+T_x^2}$ \cite{horodecki1995violating}.
\end{proposition}
\begin{proof}
Symmetric assignment errors multiply each correlator by $c_Ac_B$, and the depolarising channel multiplies Bob's Bloch vector by $1-\eps_T$. Inserting the settings into Eq.~\eqref{eq:chsh} gives $E_{00}+E_{01}=2(1-\eps_T)a\cos\theta$ and $E_{10}-E_{11}=2(1-\eps_T)b\sin\theta$. Maximising $a\cos\theta+b\sin\theta$ over $\theta$ yields $\sqrt{a^2+b^2}$ at $\tan\theta^\star=b/a$. The Horodecki criterion states that the maximal CHSH value of a two-qubit state equals twice the square root of the sum of the two largest eigenvalues of $T^{\mathsf T}T$, which for a diagonal $T$ with $|T_z|,|T_x|\ge|T_y|$ is $2\sqrt{T_z^2+T_x^2}$.
\end{proof}

\begin{corollary}
\label{cor:aniso}
If the longitudinal correlation is preserved, $a\to1$, and the transverse correlation is damped to $b\ll1$, then $S^\star\to2\sqrt{1+b^2}>2$ for every $b>0$, whereas the symmetric settings give $S(\pi/4)=\sqrt2(1+b)$, which exceeds 2 only for $b>\sqrt2-1\approx0.414$. Anisotropic memories therefore retain a certified violation, though a vanishing one, at settings adapted to the anisotropy.
\end{corollary}

The practical size of the effect follows from Eq.~\eqref{eq:Stheta}. For $a=0.98$ and $b=0.5$, the symmetric settings give $S=2.093$ and the adapted settings $S^\star=2.200$; at $Q=\SI{1}{\percent}$ the asymptotic rate rises from zero to 0.077 bit per round. Since $\theta^\star$ is in general not a multiple of $\pi/4$, a tunable non-Clifford rotation is more valuable than a fixed $T$ gate. The angle must be fixed before the protocol starts, from a separate calibration run; choosing it from the statistics of the same run would make the settings depend on the data and is not covered by the security proof. A mis-estimated $\theta$ lowers $S^\star$ by $\cos(\theta-\theta^\star)$ and cannot compromise security. This supports the first part of Hypothesis~\ref{hyp:aniso}; the second part is tested in Sec.~\ref{sec:res_qs}.

\subsection{Clifford-Native Alternative: the Magic Square}
\label{sec:ms}
In the Mermin-Peres magic-square game \cite{mermin1990simple,peres1990incompatible}, Alice receives a row and Bob a column of the array
\begin{equation}
\begin{pmatrix} X\mathbb{1} & \mathbb{1}X & XX\\ \mathbb{1}Y & Y\mathbb{1} & YY\\ XY & YX & ZZ \end{pmatrix},
\label{eq:ms}
\end{equation}
whose entries commute within each row and each column. They return $\pm1$ values whose product is $+1$ along rows and $(+1,+1,-1)$ along columns, and they win when their values agree on the shared entry. Two Bell pairs allow a perfect strategy, while the best classical strategy wins with probability $8/9$. With two tetrons per node, every entry of Eq.~\eqref{eq:ms} is a product of two bilinears, hence a four-Majorana parity: for instance $X_LX_L=-\gamma_2\gamma_3\gamma_6\gamma_7$. Joint parities of four Majoranas across two islands are the primitive operation of measurement-based Majorana architectures \cite{karzig2017scalable,bonderson2008measurement}, so the game needs no non-Clifford resource. Any certified key requires a winning probability above the local bound $8/9$. The DI-QKD protocol built on this game \cite{zhen2023device} gives a positive asymptotic key against collective attacks for winning probabilities above $0.9575$ in a white-noise model; because the key rate depends on the full statistics and not only on $\omega_{\mathrm{MS}}$, we use this value as an indicative benchmark rather than as a threshold for our noise model. Since each round consumes two pairs, the protocol yields at most half a key bit per pair even in the ideal case.
\begin{algorithm*}[!t]
\caption{Block-based flagged heralded DI-QKD with Majorana memories}
\label{alg:protocol}
\begin{algorithmic}[1]
\Require number of blocks $m$, test probability $\mu$, maximal block length $s_{\max}=\lceil1/\mu\rceil$, expected winning probability $\omega_{\mathrm{exp}}$, tolerance $\delta_{\mathrm{est}}$, setting angle $\theta$ (fixed by a separate calibration)
\For{$j=1,\dots,m$} \Comment{block $j$}
  \State $s\gets0$, $C_j\gets\perp$
  \Repeat
    \State $s\gets s+1$
    \Repeat \Comment{attempt loop; no input exists yet}
      \State both nodes initialise their tetrons and emit (two emission rounds for Barrett-Kok)
      \State the station announces the click pattern $h$
      \State if $h$ is a valid herald, each node measures its total island parity and records $F_A$, $F_B$
      \State the nodes commit $(h,F_A,F_B)$ to the public log
    \Until{$h$ is a valid herald and $F_A=F_B=0$}
    \State draw the round type $T_{j,s}\sim\mathrm{Bernoulli}(\mu)$ from shared trusted randomness
    \If{$T_{j,s}=0$}
      \State Alice sets $x=0$ and Bob sets $y=2$ \Comment{key round, Clifford only}
    \Else
      \State Alice draws $x\in_R\{0,1\}$ and Bob draws $y\in_R\{0,1\}$ from local trusted RNGs \Comment{test round}
    \EndIf
    \State measure according to Eq.~\eqref{eq:bob}; a readout that fails to converge is assigned $+1$
  \Until{$T_{j,s}=1$ or $s=s_{\max}$} \Comment{a block ends at its first test round}
\EndFor
\State announce the inputs and outputs of all test rounds; for each block with a test round set $C_j=1$ if the CHSH game was won and $C_j=0$ otherwise
\State let $N$ be the total number of rounds and $W=|\{j:C_j=1\}|$; abort if $W<m\,p_{\mathrm{blk}}(\omega_{\mathrm{exp}}-\delta_{\mathrm{est}})$, where $p_{\mathrm{blk}}=1-(1-\mu)^{s_{\max}}$
\State perform error correction on the $N$ raw bits with leakage $\mathrm{leak}_{\mathrm{EC}}$, verify with a hash, and apply privacy amplification to the length $\ell$ of Eq.~\eqref{eq:ell}
\end{algorithmic}
\end{algorithm*}

The cost is memory. The first pair must be stored while the second is being heralded, for an average time $\bar T_w=1/(R_{\mathrm{att}}P_{\mathrm{kept}})$. We use the following noise model: independent assignment errors $\eps_2$ on each of the two measured four-Majorana parities of a row or column, with the third entry inferred from the product, and memory damping of the stored pair averaged over an exponentially distributed waiting time. Sec.~\ref{sec:res_ms} evaluates the resulting trade-off against the non-Clifford route.

\section{Protocol}
\label{sec:protocol}

Algorithm~\ref{alg:protocol} states the protocol. Rounds are grouped into blocks of at most $s_{\max}=\lceil1/\mu\rceil$ rounds, each block ending at its first test round, which is the structure required by the finite-size bound of Sec.~\ref{sec:security} \cite{arnon2018practical,murta2019towards}. The algorithm implements this structure explicitly. The number of blocks $m$ is fixed in advance, so the total number of rounds $N$ is random, with expectation $\bar n=m\bar s$ where $\bar s=p_{\mathrm{blk}}/\mu$ is the expected block length. Each block contributes one statistics symbol $C_j\in\{1,0,\perp\}$ (won test, lost test, no test), and parameter estimation compares the number of won blocks with $m\,p_{\mathrm{blk}}(\omega_{\mathrm{exp}}-\delta_{\mathrm{est}})$, which is the event on which the entropy-accumulation bound is conditioned. A round inside a block is a kept round: the attempt loop that precedes it involves no input and is treated in Lemma~\ref{lem:flag}.

\subsection{Just-in-Time Parity Flag}
\begin{lemma}[Soundness of flag post-selection]
\label{lem:flag}
Let $K_i\in\{0,1\}$ decide whether attempt $i$ is kept. If $K_i$ is a function of the herald and of the two flags only, and is committed before the inputs $(X_i,Y_i,T_i)$ are drawn from trusted RNGs that are independent of all previous registers, then the kept rounds satisfy the assumptions of the DI security proof of \cite{arnon2018practical}, and the key-length bound of Sec.~\ref{sec:security} applies to them unchanged, with the block structure of Algorithm~\ref{alg:protocol} defined on the kept rounds.
\end{lemma}
\begin{proof}
Define the channel of kept round $i$ to comprise all attempts after kept round $i-1$: the devices emit, the station announces $h$, the flags are produced, and the loop repeats until $K=1$; only then are the inputs of round $i$ drawn and the measurements made. The herald and the flags of every attempt are published and are included in the side information held by Eve. This defines a sequence of channels acting on the devices' internal memory, exactly as in \cite{arnon2018practical}, in which the number of discarded attempts is an internal variable of the channel. The inputs of round $i$ are drawn from trusted RNGs after the channel has fixed every register of the attempt loop, so they are uniform and independent of all earlier registers, and the Markov conditions of the EAT hold with the published heralds and flags placed in the conditioning system. The number of blocks $m$ is fixed in advance. The attempt loop stops on public data produced before any input, and each block of Algorithm~\ref{alg:protocol} ends on the round-type bit $T_{j,s}$, which is drawn from trusted randomness and is part of the input register of the block protocol of \cite{arnon2018practical}; neither stopping rule therefore introduces a dependence between inputs and outputs beyond those that the block-wise entropy accumulation already allows. The event-ready herald of loophole-free Bell tests \cite{hensen2015loophole} is the special case without flags. Since the flag is produced by untrusted hardware, an adversarial flag can only reduce the number of kept rounds or degrade the observed statistics, and both effects are caught by the abort test.
\end{proof}

The flag is a single measurement of the total island parity, which commutes with all logical operators by Eq.~\eqref{eq:logical}, taken after the herald has arrived and before the inputs exist. By Lemma~\ref{lem:poison} it detects every poisoning event that changed the island parity during $\tpre$, which is the dominant window at long distance. Events during $\tpost$, along with all intrinsic switching and quasi-static dephasing, are not flagged and enter $S$ and $Q$ through Eq.~\eqref{eq:memory}. The flag carries two requirements. The flag must be committed before the inputs are drawn, which is a timing condition inside each laboratory rather than a relativistic one. The parity measurement must be faster than $1/\Gqp$ and must not reveal the logical state. In a floating, Coulomb-blockaded island the total parity is fixed by the island charge, so a charge measurement reads it without coupling to the bilinears \cite{plugge2017majorana,karzig2017scalable}. In a grounded device the two parity states can be charge neutral and indistinguishable to a probe that does not couple the modes, as observed in a minimal Kitaev chain \cite{vanloo2026single}, so the flag requires the floating-island geometry of the tetron.

Selections that are not of this form are excluded. Discarding blocks whose observed statistics look poor, adapting the test fraction to the observed variance of $S$, or rejecting rounds on the basis of a flag measured after the inputs are known are all outside Lemma~\ref{lem:flag}, and each can be exploited by an adversarial device. We use none of them.

\subsection{Detection Loophole in a Heralded Architecture}
\begin{corollary}[Local readout efficiency]
\label{cor:eff}
Let a readout converge with probability $\eta_r$, with failed readouts assigned $+1$, and let the assignment errors be $\eps_0=P(-|+)$ and $\eps_1=P(+|-)$. Each outcome then satisfies $\mathbb{E}[a'|a]=c\,a+d$ with $c=\eta_r(1-\eps_0-\eps_1)$ and $d=\eta_r(\eps_1-\eps_0)+1-\eta_r$, and for states with vanishing marginals the observed correlators are $E'_{xy}=c_Ac_BE_{xy}+d_Ad_B$. For an ideal Bell pair with symmetric $\eta_r$ and no assignment error this gives $S=2\sqrt2\eta_r^2+2(1-\eta_r)^2$, which exceeds 2 if and only if $\eta_r>2/(1+\sqrt2)\approx0.828$.
\end{corollary}
\begin{proof}
The map $\mathbb{E}[a'|a]=ca+d$ follows from evaluating the two conditional expectations. For a product of independent local maps, $\mathbb{E}[a'b']=c_Ac_B\mathbb{E}[ab]+c_Ad_B\mathbb{E}[a]+d_Ac_B\mathbb{E}[b]+d_Ad_B$, and the two middle terms vanish when the marginals do. Substituting the ideal correlators $E_{xy}=\pm1/\sqrt2$ into Eq.~\eqref{eq:chsh} gives the stated $S$, and solving $S=2$ gives $\eta_r=2/(1+\sqrt2)$.
\end{proof}

The threshold coincides with the value derived by Garg and Mermin for a direct Bell test \cite{garg1987detector}, but the quantity it constrains is different: $\eta_r$ is the convergence probability of the local parity readout, which dispersive charge sensing achieves with a margin \cite{microsoft2025interferometric,vanloo2026single}. Photon loss enters only through the rate. Monitoring setting-conditional detection efficiencies and subtracting a heuristic penalty from $S$, a procedure sometimes proposed for heralded architectures, is therefore neither necessary nor sufficient; binning the failed readouts is the treatment that carries a proof.

\section{Security Analysis and Key Length}
\label{sec:security}

\subsection{Security Definitions}
A protocol is $\eps^{\mathrm c}$-complete if honest devices abort with probability at most $\eps^{\mathrm c}$, and $\eps^{\mathrm s}$-sound if the joint state of the output key, the abort flag, the public transcript and Eve's quantum side information is $\eps^{\mathrm s}$-close in trace distance to an ideal state in which the key is uniform, identical for both parties and independent of Eve \cite{portmann2022security}. Soundness composes as $\eps^{\mathrm s}\le\eps_{\mathrm{EA}}+\eps_{\mathrm{EC}}+\eps_{\mathrm{PA}}+\eps_{s}$, with the four terms covering entropy accumulation, error-correction failure, privacy amplification and smoothing \cite{arnon2018practical}. We take $\eps^{\mathrm s}=10^{-8}$ and split it equally, and we set the completeness parameters to $\eps^{\mathrm c}_{\mathrm{EA}}=\eps'_{\mathrm{EC}}=10^{-2}$, so that honest devices abort with probability at most $\eps^{\mathrm c}\le2\times10^{-2}$; the number of blocks is fixed in advance, so fluctuations of the flag and herald rates affect only the duration of the run, and fluctuations of the block lengths affect only the realised number of rounds $N$. The authentication of the classical channel consumes a short pre-shared key and contributes an additional additive term that we do not repeat here.

\subsection{Min-Tradeoff Function}
Write the single-round entropy bound of Eq.~\eqref{eq:pironio} as a function of the winning probability,
\begin{equation}
g(\omega)=1-h\!\left(\tfrac12+\tfrac12\sqrt{16\,\omega(\omega-1)+3}\right),
\label{eq:g}
\end{equation}
valid for $\omega\in[3/4,(2+\sqrt2)/4]$, which follows from Eq.~\eqref{eq:pironio} through $S=8\omega-4$. The tangents of Eq.~\eqref{eq:g} are the affine functions $f$ that enter Eq.~\eqref{eq:eatv}.

\begin{lemma}[Convexity]
\label{lem:convex}
$g$ is convex and increasing on $[3/4,(2+\sqrt2)/4]$, so its tangent at any interior point $\omega_t$ lower-bounds it on the whole interval.
\end{lemma}
\begin{proof}
Write $g=\varphi(u)$ with $\varphi(u)=1-h(\tfrac12+\tfrac12u)$ and $u=\sqrt{S^2/4-1}$, where $S=8\omega-4$ is affine in $\omega$, so it suffices to prove convexity in $S$ on $(2,2\sqrt2]$. One has $\varphi'(u)=\tfrac12\log_2\frac{1+u}{1-u}\ge0$ and $\varphi''(u)=\frac{1}{\ln2}\frac{1}{1-u^2}>0$, while $u'=S/(4u)>0$ and $u''=-1/(4u^3)<0$. Hence
\[
8u^3\ln 2\;g''(S)=\frac{2u(1+u^2)}{1-u^2}-\ln\frac{1+u}{1-u}.
\]
Expanding both terms for $u\in(0,1)$ gives $2u+4u^3+4u^5+\dots$ and $2u+\tfrac23u^3+\tfrac25u^5+\dots$, so the difference is positive term by term and $g''>0$. Monotonicity follows from $\varphi'\ge0$ and $u'>0$. A convex function lies above each of its tangents.
\end{proof}

By Lemma~\ref{lem:convex} the min-tradeoff function can be taken to be the tangent of $g$ at a point $\omega_t$ that we optimise, expressed in terms of the frequency of winning test rounds. Since a test round occurs with probability $\mu$, the gradient with respect to that frequency is $g'(\omega_t)/\mu$, which is what enters the variance term of Eq.~\eqref{eq:eat}.

\subsection{Key Length}
\begin{theorem}[Finite key length, after \cite{arnon2018practical,murta2019towards}]
\label{thm:key}
Consider Algorithm~\ref{alg:protocol} with $m$ blocks, test probability $\mu$, $s_{\max}=\lceil1/\mu\rceil$, $p_{\mathrm{blk}}=1-(1-\mu)^{s_{\max}}$ and expected block length $\bar s=p_{\mathrm{blk}}/\mu$, and let $n$ denote the number of rounds that enters error correction and the estimation of Bob's test outputs, namely the realised number $N$ in the protocol and its expectation $\bar n=m\bar s$ in the numerical evaluation. Let $\delta_{\mathrm{est}}=\sqrt{2\omega_{\mathrm{exp}}\ln(1/\eps^{\mathrm c}_{\mathrm{EA}})/(m\,p_{\mathrm{blk}})}$, and define
\begin{multline}
\eta_{\mathrm{opt}}=\max_{\omega_t\in\mathcal{W}}\Big\{\bar s\big[g(\omega_t)+g'(\omega_t)(\omega_{\mathrm{exp}}-\delta_{\mathrm{est}}-\omega_t)\big]\\
-\tfrac{2}{\sqrt m}\big[\log_2(1+2\cdot6^{s_{\max}})+\lceil g'(\omega_t)/\mu\rceil\big]\\
\times\sqrt{1-2\log_2[\eps_s(\eps_{\mathrm{EA}}+\eps_{\mathrm{EC}})]}\Big\},\quad \mathcal{W}=\big(\tfrac34,\tfrac{2+\sqrt2}{4}\big),
\label{eq:etaopt}
\end{multline}
\begin{multline}
\mathrm{leak}_{\mathrm{EC}}= n\big[(1-\mu)h(Q)+\mu h(\omega_{\mathrm{exp}})\big]\\
+4\sqrt n\log_2(2\sqrt2+1)\sqrt{\log_2\tfrac{8}{\eps_{\mathrm{EC}}'^2}}\\
+\log_2\!\Big(\tfrac{8}{\eps_{\mathrm{EC}}'^2}+\tfrac{2}{2-\eps'_{\mathrm{EC}}}\Big)+\log_2\tfrac{1}{\eps_{\mathrm{EC}}}.
\label{eq:leak}
\end{multline}
Then, by the entropy accumulation argument of \cite{arnon2018practical} applied block by block to the kept rounds of Algorithm~\ref{alg:protocol}, the protocol is $\eps^{\mathrm s}$-sound and $\eps^{\mathrm c}$-complete with an extractable key of length
\begin{multline}
\ell=m\,\eta_{\mathrm{opt}}-\mathrm{leak}_{\mathrm{EC}}-3\log_2\!\big(1-\sqrt{1-\eps_s^2}\big)-\mu n\\
-2\log_2(7)\sqrt n\sqrt{1-2\log_2\eps_s}-2\log_2\tfrac{1}{\eps_{\mathrm{PA}}}.
\label{eq:ell}
\end{multline}
\end{theorem}
The entropy-accumulation term $m\,\eta_{\mathrm{opt}}$ is accumulated over the $m$ blocks, each block acting as one step of the theorem with output register of dimension at most $6^{s_{\max}}$ and statistics register $C_j$; this is why the protocol must fix $m$ and must end each block at its first test round or after $s_{\max}$ rounds, as Algorithm~\ref{alg:protocol} does. The realised $N$ is public before error correction, so the parties evaluate Eqs.~\eqref{eq:leak} and \eqref{eq:ell} with $n=N$; since $N$ concentrates around $\bar n$ with relative fluctuations of order $m^{-1/2}$, we evaluate the numerics at $n=\bar n$. We do not claim this theorem as a new result. It restates the block-protocol key length of \cite[Methods]{arnon2018practical}, in the form used by \cite{murta2019towards}, with constants that are upper bounds on the ones appearing there; since every such replacement enlarges a penalty term, the resulting $\ell$ is a valid, if pessimistic, key length. The constants are the following; each item either reproduces a term of that bound or replaces it by an upper bound on the corresponding penalty, so that the key length can only decrease.
\begin{itemize}
\item The output register of one block consists of Alice's $s_{\max}$ outcomes and one statistics symbol in $\{\text{win},\text{lose},\perp\}$, of dimension $3\cdot2^{s_{\max}}$. We use the larger value $6^{s_{\max}}$, which would also allow one symbol per round, and bound $\log_2(1+2\cdot6^{s_{\max}})\le1+s_{\max}\log_26$. The validation below shows that the tighter dimension lowers $n_{\min}$ by about \SI{20}{\percent}.
\item The gradient of the min-tradeoff function with respect to the block statistics is $g'(\omega_t)/\mu$, and we round it up as required by Eq.~\eqref{eq:eat}.
\item The smooth max-entropy of Bob's published test outputs is bounded by one bit per test round plus the second-order term of Eq.~\eqref{eq:ell}, which is the form used in \cite{arnon2018practical}.
\item The abort threshold $\omega_{\mathrm{exp}}-\delta_{\mathrm{est}}$ is a protocol parameter: soundness holds for any value, and we choose $\delta_{\mathrm{est}}$ from a multiplicative Chernoff bound on the number of won test rounds so that honest devices pass with probability at least $1-\eps^{\mathrm c}_{\mathrm{EA}}$.
\item The security parameters are split evenly, $\eps_{\mathrm{EA}}=\eps_{\mathrm{EC}}=\eps_{\mathrm{PA}}=\eps_s=\eps^{\mathrm s}/4$.
\end{itemize}
 The terms $\mu n$ and $2\log_2(7)\sqrt n\sqrt{\cdot}$ bound the smooth max-entropy of Bob's test outputs, which are published during parameter estimation. Lemma~\ref{lem:flag} guarantees that the theorem applies to the kept rounds of Algorithm~\ref{alg:protocol}.

\emph{Validation and sensitivity.} With the nitrogen-vacancy parameters $S=2.47$ and $Q=\SI{5.1}{\percent}$ and a soundness of $10^{-5}$, our implementation requires $n_{\min}=3.3\times10^8$ rounds, against of order $10^8$ read from Fig.~4 of \cite{murta2019towards}. Replacing the output-alphabet bound $6^{s_{\max}}$ by the tighter $3\cdot2^{s_{\max}}$ lowers the requirement to $2.5\times10^8$, and lowering the completeness tolerance lowers it further to $2.2\times10^8$; the remaining factor of about two is within the accuracy with which the reference value can be read from a logarithmic plot and the differences in how the two implementations allocate the error parameters. The discrepancy is on the conservative side in every case. For the operating points of Sec.~\ref{sec:results} the tighter alphabet reduces $n_{\min}$ by \SI{20}{\percent} to \SI{22}{\percent}, which changes none of the conclusions, so we keep the conservative constant throughout.

\subsection{Scaling of the Finite-Size Penalty}
\begin{proposition}[Optimal test fraction]
\label{prop:scaling}
Let $\Delta(n)=r_\infty-\ell/n$ be the finite-size penalty of Theorem~\ref{thm:key} at fixed $(S,Q)$, with $n=\bar n$ the expected number of rounds. In the regime where the statistical correction is small compared with $r_\infty$, so that the $\mu$-dependent part of the penalty consists of the direct cost of testing and the second-order term of Eq.~\eqref{eq:etaopt}, the optimum test fraction and the penalty obey
\begin{equation}
\mu^\star\propto n^{-1/3},\qquad \Delta(n)\propto n^{-1/3}.
\label{eq:scaling}
\end{equation}
\end{proposition}
\begin{proof}
Two groups of terms in Eqs.~\eqref{eq:etaopt} and \eqref{eq:ell} depend on $\mu$. The direct cost of testing, namely the term $\mu n$ together with the difference between $h(Q)$ and $h(\omega_{\mathrm{exp}})$ in Eq.~\eqref{eq:leak}, contributes $\kappa\mu$ per round with $\kappa=1+h(\omega_{\mathrm{exp}})-h(Q)$. The second-order term of Eq.~\eqref{eq:etaopt} contributes $2\big[\log_2(1+2\cdot6^{s_{\max}})+g'(\omega_t)/\mu\big]\sqrt{\cdot}/(\bar s\sqrt m)$ per round; both bracketed terms grow as $1/\mu$, while $\bar s\sqrt m=\sqrt{\bar sn}$ with $\bar s\to(1-e^{-1})/\mu$, so this group behaves as $B\mu^{-1/2}n^{-1/2}$. Hence $\Delta(\mu,n)\simeq\kappa\mu+B\mu^{-1/2}n^{-1/2}$ up to terms that do not depend on $\mu$. Setting $\partial\Delta/\partial\mu=0$ gives $\mu^\star=[B/(2\kappa)]^{2/3}n^{-1/3}$ and $\Delta(n)\simeq3\kappa\mu^\star\propto n^{-1/3}$.
\end{proof}

Numerically, fitting the optimiser output for $n\ge10^9$ at the Tier~II operating point of Sec.~\ref{sec:results} gives $\mu^\star\propto n^{-0.334}$ and $\Delta\propto n^{-0.330}$, in agreement with Eq.~\eqref{eq:scaling}. At that point $\kappa=1.60$, so the proposition predicts $\Delta/\mu^\star\simeq4.8$; the computed ratio runs from 4.4 at $n=10^{10}$ to 5.3 at $n=10^{14}$, the drift coming from the second-order terms that do not depend on $\mu$. The practical reading is that the number of rounds needed to reach a fixed fraction of the asymptotic rate grows as the cube of the inverse of the tolerated penalty, which is the reason why the round requirements in Sec.~\ref{sec:results} are large even at high $S$.

\subsection{Residual Assumptions and Attack Surface}
Three points deserve emphasis. First, DI security removes assumptions about the internal physics of the devices, not about the laboratories: a covert channel that leaks the raw outcomes defeats any DI protocol, so the cryogenic wiring and the readout electronics must be treated as part of the secure perimeter. Second, the devices are assumed to be used sequentially and not to be re-used across protocol executions unless the security parameters are recomposed. Third, the RNGs must be independent of the devices; seeded expansion from a short private seed is compatible with the EAT but changes the accounting. None of these assumptions is specific to Majorana hardware, but the first is more demanding here than for atomic systems, because the memories sit inside a cryostat that also contains amplifiers and control lines.

\section{Closed-Form Bell Statistics and Error Budget}
\label{sec:closedform}

\begin{proposition}[Closed-form statistics]
\label{prop:closed}
Under the models of Sec.~\ref{sec:system}, with both nodes subject to the same timing,
\begin{align}
T_z&=t_{zz}f_z^2,\qquad T_x=t_{xx}f_x^2,\label{eq:T}\\
Q&=\tfrac12\big(1-c_Z^2T_z\big),\label{eq:Q}\\
S^\star&=2(1-\eps_T)\,c_Z\sqrt{c_Z^2T_z^2+c_X^2T_x^2},\label{eq:Sstar}
\end{align}
where $t_{zz}$ and $t_{xx}$ come from Eq.~\eqref{eq:bk} or \eqref{eq:spi} and $f_{z,x}$ from Eq.~\eqref{eq:memory}. An attempt yields a kept round with probability $P_{\mathrm{kept}}=P_hP_{\mathrm{even}}^2$ when the flag is used and $P_h$ otherwise, and the asymptotic key rate in bits per second is $K=M R_{\mathrm{att}}P_{\mathrm{kept}}\,r_\infty(S^\star,Q)$.
\end{proposition}
\begin{proof}
The channels of Eq.~\eqref{eq:memory} act independently at the two nodes and multiply each diagonal correlation by the product of the two per-node factors. Proposition~\ref{prop:adapted} then gives Eq.~\eqref{eq:Sstar} with $a=c_Z^2T_z$ and $b=c_Xc_ZT_x$, and Eq.~\eqref{eq:Q} follows from $E_{02}=c_Z^2T_z$ because key rounds use $Z_L$ at both nodes and contain no rotation. The factor $P_{\mathrm{even}}^2$ of Lemma~\ref{lem:flagdamp} is the probability that neither node raised a flag.
\end{proof}

We verified Eqs.~\eqref{eq:T}-\eqref{eq:Sstar} against an exact evaluation of Eq.~\eqref{eq:chsh} with two-qubit density matrices, Pauli memory channels and noisy positive-operator-valued measures, for every parameter tier, both heralding schemes and distances from 0 to \SI{300}{km}; the two agree to $10^{-9}$.

\subsection{Linearised Error Budget}
Expanding Eqs.~\eqref{eq:Q} and \eqref{eq:Sstar} to first order in all error parameters, with the flag active and $\Gint=\Gamma^{\mathrm{int}}_x+\Gamma^{\mathrm{int}}_z$,
\begin{align}
\frac{S^\star}{2\sqrt2}&\simeq1-\eps_T-3\eps_Z-\eps_X-\tfrac32\zeta-\tfrac{1-V}{2}\nonumber\\
&\quad-2\Gqp\tpost-(\Gqp\tpre)^2-2\Gint t-\tfrac12\sigma_\delta^2t^2,\label{eq:linS}\\
Q&\simeq2\eps_Z+\zeta+\Gqp\tpost+\tfrac12(\Gqp\tpre)^2+2\Gamma^{\mathrm{int}}_x t.\label{eq:linQ}
\end{align}
Here the poisoning terms of Eqs.~\eqref{eq:linS} and \eqref{eq:linQ} are written for uniform weights and an active flag, so that the pre-flag window contributes only at second order through Lemma~\ref{lem:flagdamp}. The sensitivities are $\partial S/\partial\eps_Z=-6\sqrt2$, $\partial S/\partial\eps_X=\partial S/\partial\eps_T=-2\sqrt2$ and $\partial S/\partial\zeta=-3\sqrt2$. The $Z$ readout is three times as damaging to $S$ as the $X$ readout, because it appears in Alice's key setting and in all of Bob's settings, and it is the only assignment error that enters $Q$. This ordering is a consequence of the protocol design of Sec.~\ref{sec:nonlocality}, not a property of the hardware, and it should guide calibration effort.

\subsection{Distance Limits of the Individual Channels}
Each noise process imposes its own distance limit, obtained by setting $r_\infty=0$ with all other errors switched off and $t\simeq L/\cfib$.

\emph{Poisoning without the flag.} For uniform weights the state is Werner-like with visibility $e^{-2\Gqp L/\cfib}$, so a key exists only for
\begin{equation}
L<L_{\mathrm{qp}}=\frac{\ln[1/(1-2Q^\star)]}{2}\,\frac{\cfib}{\Gqp}\approx0.077\,\frac{\cfib}{\Gqp},
\label{eq:Lqp}
\end{equation}
with $Q^\star=\SI{7.15}{\percent}$. This gives \SI{154}{km} at $\Gqp=\SI{100}{s^{-1}}$ and \SI{15}{km} at $\Gqp=\SI{1000}{s^{-1}}$. With the flag the same process reduces the rate by $P_{\mathrm{even}}^2$ and degrades the correlations only at second order, $\Phi(1/2,\tpre)^2\simeq1-(\Gqp L/\cfib)^2$, so the distance limit is pushed out by roughly the square root of the same ratio and disappears entirely when poisoning acts on a single mode.

\emph{Intrinsic switching.} For isotropic Markovian switching the visibility is $e^{-4\Gamma^{\mathrm{int}}L/\cfib}$ and the limit is $L_{\mathrm{int}}\approx0.077\,\cfib/\Gint$, which cannot be removed by flagging because the total parity is unchanged.
\begin{figure*}[!t]
\centering
\includegraphics[width=0.92\textwidth]{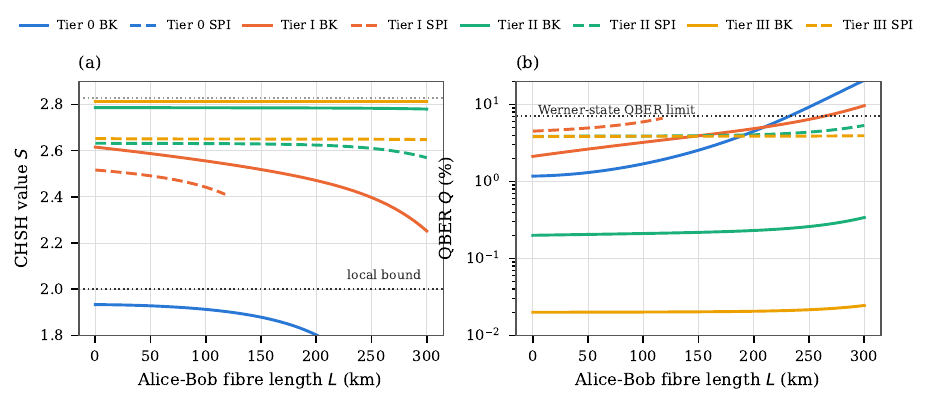}
\caption{(a) CHSH value $S^\star$ and (b) QBER against total fibre length for two-photon Barrett-Kok heralding (solid) and single-photon interference with optimised bright-state population (dashed). Tier~0, under the Markovian reading of the prototype data, violates no Bell inequality at any distance, and its single-photon curve lies below the plotted range. Single-photon curves are drawn only where the key rate is positive. The dotted line in (b) marks the Werner-state QBER limit of Eq.~\eqref{eq:pironio}.}
\label{fig:SQ}
\end{figure*}

\emph{Quasi-static splitting.} Here the longitudinal correlation is untouched, so $Q$ does not grow, and by Corollary~\ref{cor:aniso} the CHSH value stays above 2 for any finite transverse correlation. A hard limit appears only in combination with a non-zero $Q$ from other sources. Sec.~\ref{sec:res_qs} evaluates the resulting requirement on $\sigma_\delta$.

\subsection{Scaling with Channel Loss}
The two heralding schemes differ in their loss exponent. Writing the total transmission as $\eta_{\mathrm{tot}}=10^{-\alpha L/10}$, two-photon heralding succeeds with a probability proportional to $\eta_{\mathrm{tot}}$, whereas single-photon heralding succeeds with a probability proportional to $\sqrt{\eta_{\mathrm{tot}}}$, at the price of the error floor $\alpha_b$ in Eq.~\eqref{eq:spi}. The key rate in bits per second carries an additional factor $R_{\mathrm{att}}\propto(L/\cfib+t_{\mathrm{ops}})^{-1}$, which steepens both curves at long distance. Fitting the computed rates between \SI{100}{km} and \SI{200}{km} gives effective exponents of $1.15$ for Barrett-Kok and $0.66$ for single-photon heralding at Tier II parameters, and $1.15$ and $0.65$ at Tier III, close to the ideal values of 1 and $1/2$ with the geometric correction. The square-root scaling is the same advantage that single-photon and twin-field schemes enjoy over direct transmission \cite{lucamarini2018overcoming,kolodynski2020device}, and it is the reason why single-photon heralding wins at long distance in Sec.~\ref{sec:results} despite its higher QBER. The repeaterless capacity of a lossy channel \cite{pirandola2017fundamental} constrains direct transmission between two users and scales linearly in the transmission; it does not apply to a configuration with memories at the end nodes and a station in the middle, and single-photon heralding attains the square-root scaling that direct transmission cannot. What limits the present architecture instead is the round-trip latency, the memory lifetime and the interface efficiency.

\section{Numerical Results}
\label{sec:results}

Unless stated otherwise the flag is active, the settings are adapted according to Proposition~\ref{prop:adapted}, and a single memory pair is used.
\begin{table}[!tb]
\centering
\caption{Parameter tiers. Tier~0 is prototype-informed, Tier~I near-term, Tier~II a target and Tier~III optimistic. Tier~0 is inferred from the prototype values of Table~\ref{tab:sota}: the poisoning rate follows from $\Gqp=2/\tau_Z$, because a poisoning event flips $Z_L$ with probability one half by Lemma~\ref{lem:poison}, and the intrinsic $X_L$ switching from $\Gamma^{\mathrm{int}}_z=1/\tau_X$. Interface parameters are hypothetical in every tier, and all values in Tiers~I to III are engineering targets rather than measured quantities; Table~\ref{tab:prov} gives the provenance of each entry. The tiers use $\sigma_\delta=0$, and the quasi-static channel is studied separately in Sec.~\ref{sec:res_qs}. Common to all tiers: $\alpha=\SI{0.2}{dB/km}$, $\cfib=\SI{2e8}{m/s}$.}
\label{tab:tiers}
\begin{tabular}{@{}lcccc@{}}
\toprule
Parameter & Tier 0 & Tier I & Tier II & Tier III\\
\midrule
$\Gqp$ (s$^{-1}$) & 161 & 100 & 1 & 0.01\\
$\Gamma^{\mathrm{int}}_x$ (s$^{-1}$) & 0 & 10 & 0.1 & $10^{-3}$\\
$\Gamma^{\mathrm{int}}_z$ (s$^{-1}$) & $6.9\times10^4$ & 10 & 0.1 & $10^{-3}$\\
$\eps_Z$ & $5\times10^{-3}$ & $10^{-2}$ & $10^{-3}$ & $10^{-4}$\\
$\eps_X$ & 0.16 & $10^{-2}$ & $10^{-3}$ & $10^{-4}$\\
$\eps_T$ & $10^{-2}$ & $10^{-2}$ & $10^{-3}$ & $10^{-4}$\\
$\etaint$ & $10^{-2}$ & 0.05 & 0.20 & 0.50\\
$\eta_{\mathrm{det}}$ & 0.90 & 0.90 & 0.90 & 0.95\\
$V$ & 0.95 & 0.95 & 0.98 & 0.99\\
$p_n$ & $10^{-6}$ & $10^{-6}$ & $10^{-7}$ & $10^{-8}$\\
$t_{\mathrm{ops}}$ & \SI{10}{\micro s} & \SI{10}{\micro s} & \SI{1}{\micro s} & \SI{1}{\micro s}\\
$\tpost$ & \SI{10}{\micro s} & \SI{10}{\micro s} & \SI{1}{\micro s} & \SI{1}{\micro s}\\
\bottomrule
\end{tabular}
\end{table}
\begin{table}[!tb]
\centering
\caption{Provenance of the model parameters.}
\label{tab:prov}
\begin{tabular}{@{}llp{3.2cm}@{}}
\toprule
Parameter & Status & Basis\\
\midrule
$\Gqp$, Tier 0 & inferred & $2/\tau_Z$ with $\tau_Z$ from \cite{aghaee2025distinct}\\
$\Gamma^{\mathrm{int}}_z$, Tier 0 & inferred & $1/\tau_X$ from \cite{aghaee2025distinct}\\
$\eps_Z,\eps_X$, Tier 0 & measured & \cite{aghaee2025distinct}\\
$\tpost$, Tier 0 & estimated & integration time of \cite{microsoft2025interferometric}\\
$\eta_{\mathrm{det}}$ & established & superconducting nanowire detectors\\
$\alpha$, $\cfib$ & established & standard telecom fibre\\
$\etaint$, $V$, $p_n$ & hypothetical & no Majorana optical interface exists \cite{contamin2021hybrid,lauk2020perspectives}\\
$\eps_T$ & hypothetical & geometric gate \cite{karzig2016universal} or injection \cite{bravyi2005universal}\\
Tiers I to III & targets & engineering design space\\
\bottomrule
\end{tabular}
\end{table}

Table~\ref{tab:prov} states the status of every parameter used in Table~\ref{tab:tiers}. No Tier~I to III value is a measured quantity; they define a design space, and the feasibility maps presented below are the operational statement of what has to be achieved.

\subsection{Prototype-Informed Parameters}
\label{sec:res_tier0}
Under the Markovian reading of the tetron data and the timing of Table~\ref{tab:tiers}, Tier~0 gives $S=1.93$ at $L=0$, below the local bound, so no key can be certified at any distance (Fig.~\ref{fig:SQ}a). The result is insensitive to the quoted uncertainties of the inputs: varying $\tau_X$ over \SI{14.5+-0.3}{\micro s} and $\tau_Z$ over \SI{12.4+-0.4}{ms} changes $S$ by less than $10^{-3}$. The limiting quantity is $\tau_X$, which is comparable to the local operation times and suppresses the transverse correlation. If $\tau_X$ is removed from the model and all other Tier~0 values are kept, $S=2.31$, $Q=\SI{1.17}{\percent}$ and $r_\infty=0.16$, and a positive rate survives to \SI{100}{km} ($r_\infty=0.11$), so the poisoning rate implied by $\tau_Z$ does not by itself prevent a key; lowering $\eps_X$ from \SI{16}{\percent} to \SI{0.5}{\percent} in addition gives $S=2.67$ and $r_\infty=0.58$ at $L=0$. Scanning $\tau_X$ at fixed Tier~0 parameters, a positive asymptotic rate requires $\tau_X\gtrsim\SI{0.22}{ms}$ at $L=0$ and $\tau_X\gtrsim\SI{0.76}{ms}$ at \SI{10}{km}, improvements by factors of roughly 15 and 53.

The conclusion depends on how the data are read, and we therefore bracket it. Shortening the local timing to its floor, an ideal link and a post-flag time equal to the reported integration time of \SI{3.7}{\micro s}, gives $S=1.99$ under the Markovian reading, still below the bound. The reported $\tau_X$ was attributed to jumps induced by residual splitting \cite{aghaee2025distinct}; if the same scale is instead read as a quasi-static splitting with $1/e$ time $\tau_X$, the Table~\ref{tab:tiers} timing still gives $S=1.93$, while the minimal timing gives $S=2.26$ and $r_\infty=0.12$ at $L=0$, falling to zero at \SI{0.7}{km}. The most favourable reading of the 2025 data thus supports a key only below one kilometre. Two further caveats apply: $\tau_X$ was extracted from continuously monitored telegraph noise and need not equal the idle memory time, and the $X$- and $Z$-loop errors were measured in separate configurations \cite{aghaee2025distinct}. The analysis identifies $\tau_X$, its functional form, and $\eps_X$ as the quantities whose independent measurement decides feasibility, which supports Hypothesis~\ref{hyp:interface}.

\subsection{Bell Statistics and Key Rate against Distance}
For Tiers~II and III the statistics are nearly independent of distance up to \SI{300}{km} (Fig.~\ref{fig:SQ}), because the flag removes the poisoning contribution and the remaining rates are small compared with $\cfib/L$. Tier~I degrades from $S=2.62$ at $L=0$ to $S=2.25$ at \SI{300}{km}; removing either its unflagged intrinsic switching, at $\Gint=\SI{20}{s^{-1}}$, or its noise-click false heralds restores about \SI{40}{\percent} of that loss. Single-photon heralding accepts a QBER of 4 to \SI{5}{\percent}, set mainly by the optimised bright-state population $\alpha_b\approx0.02$ to $0.05$, in exchange for a success probability that scales more favourably with distance.
\begin{figure*}[!t]
\centering
\includegraphics[width=0.92\textwidth]{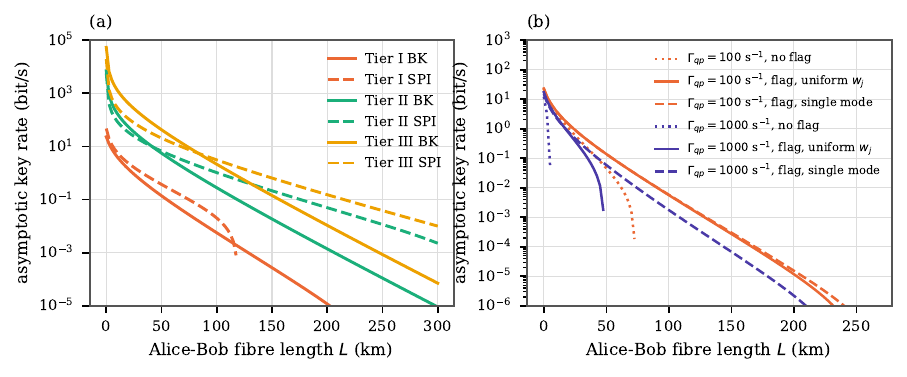}
\caption{(a) Asymptotic secret-key rate per memory pair against distance for Tiers~I to III, with Barrett-Kok heralding (solid) and single-photon interference (dashed). (b) Effect of the just-in-time parity flag for Tier~I parameters with Barrett-Kok heralding and two poisoning rates. Without the flag (dotted) poisoning imposes a hard cutoff consistent with Eq.~\eqref{eq:Lqp}. With the flag it becomes a rate penalty plus the residual second-order loss of Eq.~\eqref{eq:phi}, shown for uniform mode weights (solid) and for poisoning dominated by one mode (dashed).}
\label{fig:rate}
\end{figure*}

Fig.~\ref{fig:rate}a shows the resulting crossover. At \SI{50}{km}, Tier~II yields \SI{5.6}{bit/s} with Barrett-Kok and \SI{6.7}{bit/s} with single-photon heralding for one memory pair, and at \SI{100}{km} the two give \SI{0.28}{bit/s} and \SI{1.0}{bit/s}. For Tier~III, Barrett-Kok is faster at \SI{50}{km}, \SI{42}{bit/s} against \SI{21}{bit/s}, while single-photon heralding wins at \SI{100}{km}, \SI{3.1}{bit/s} against \SI{2.1}{bit/s}. Tier~I stops producing key with single-photon heralding near \SI{118}{km}, where noise clicks and intrinsic switching push its QBER above the threshold. The fitted loss exponents quoted in Sec.~\ref{sec:closedform} confirm that the crossover originates in the scaling of the heralding probability rather than in the fidelity. With equal visibilities the crossover lies at \SI{43}{km} for Tier~II and \SI{85}{km} for Tier~III. Because single-photon heralding also needs optical phase stability, we repeated the comparison with a penalty $V_\phi<1$: for $V_\phi=0.97$ and $0.95$ the crossover moves to \SI{50}{km} and \SI{55}{km} for Tier~II and to \SI{90}{km} and \SI{95}{km} for Tier~III, and the single-photon rate at \SI{100}{km} drops by \SI{14}{\percent} and \SI{22}{\percent}. The qualitative ordering of the two schemes is therefore robust, while the crossover distance depends on phase stabilisation at the level of tens of kilometres.

Noisy preprocessing \cite{ho2020noisy} adds little at these operating points, as expected from the low QBER: at \SI{50}{km} it raises the Tier~I single-photon rate from 0.161 to 0.166 bit per round, an improvement of \SI{2.6}{\percent}, and changes the Barrett-Kok rates by less than $10^{-4}$ bit per round. The technique becomes relevant only when the QBER approaches the \SI{7}{\percent} region, which in this architecture happens either at large $\alpha_b$ or close to the distance limits.

\subsection{Poisoning Flag}
\label{sec:res_flag}
Fig.~\ref{fig:rate}b isolates the flag for Tier~I with Barrett-Kok heralding, for the two microscopic scenarios of Lemma~\ref{lem:flagdamp}. At $\Gqp=\SI{100}{s^{-1}}$ the largest distance with a positive rate grows from \SI{72}{km} without the flag to \SI{248}{km} with the flag and uniform mode weights, and to \SI{265}{km} when poisoning is dominated by one Majorana mode. At $\Gqp=\SI{1000}{s^{-1}}$, which corresponds to the millisecond parity lifetimes reported for current devices \cite{microsoft2025interferometric,vanloo2026single}, the three cases give \SI{5}{km}, \SI{48}{km} and \SI{233}{km}. The gap between the last two numbers is the residual quadratic term of Eq.~\eqref{eq:phi}: with uniform weights a pair of poisoning events leaves the parity unchanged but scrambles the qubit, whereas two events on the same mode cancel. Which case applies is a property of the device that can be measured, by comparing the decay of flagged and unflagged rounds, and at high poisoning rates it is worth an order of magnitude in reach.

The unflagged values lie below the idealised $L_{\mathrm{qp}}$ of Eq.~\eqref{eq:Lqp}, which is \SI{154}{km} and \SI{15}{km} for the two rates, because the remaining Tier~I errors consume part of the noise budget. Poisoning, usually presented as the fundamental distance limit of the platform, is therefore removable by protocol design provided that a total-parity measurement much faster than $1/\Gqp$ is available, which supports Hypothesis~\ref{hyp:flag}. What remains are the unflagged channels and the loss.
\begin{figure*}[!t]
\centering
\includegraphics[width=0.96\textwidth]{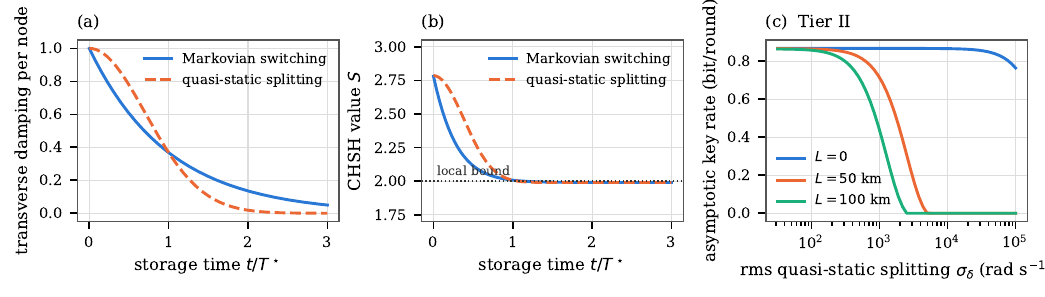}
\caption{Markovian against quasi-static transverse noise at equal $1/e$ time $T^\star$. (a) Damping of the transverse correlation per node. (b) Resulting CHSH value with Tier~II readout and $L=0$. (c) Asymptotic key rate against the rms quasi-static splitting $\sigma_\delta$ for three distances at Tier~II parameters.}
\label{fig:qs}
\end{figure*}

Imperfect flags change these numbers smoothly, as Eq.~\eqref{eq:phiimpexp} predicts. At $\Gqp=\SI{100}{s^{-1}}$ with uniform weights, miss probabilities $\eps_m$ of \SI{1}{\percent}, \SI{3}{\percent} and \SI{10}{\percent} reduce the reach from \SI{248}{km} to \SI{245}{km}, \SI{240}{km} and \SI{218}{km}. At $\Gqp=\SI{1000}{s^{-1}}$ the uniform-weight reach falls from \SI{48}{km} to about \SI{45}{km}, \SI{43}{km} and \SI{33}{km}, while the single-mode reach, which relies on the complete cancellation of the first-order term, falls from \SI{233}{km} to \SI{185}{km}, \SI{108}{km} and \SI{38}{km}. A false-alarm probability of \SI{5}{\percent} leaves $S$ unchanged and lowers the rate by \SI{10}{\percent}. The flag thus retains most of its benefit with miss probabilities of a few per cent, but the order-of-magnitude gain of single-mode poisoning at high rates requires $\eps_m\lesssim\SI{1}{\percent}$, which is a concrete specification for the total-parity readout.

\subsection{Noise Anisotropy and the Form of the Decay}
\label{sec:res_qs}
Fig.~\ref{fig:qs} compares a Markovian switching process and a quasi-static splitting whose transverse damping has the same $1/e$ time $T^\star$. At $t=0.1\,T^\star$ the Markovian model leaves a damping factor of 0.909 and the quasi-static model 0.991, which translates into $S=2.56$ against $S=2.76$; at $t=0.3\,T^\star$ the factors are 0.739 and 0.912 and the CHSH values 2.26 and 2.57. The difference is the quadratic onset $1-\sigma_\delta^2t^2/2$ of Gaussian dephasing against the linear onset $1-2\Gamma t$ of a Poisson process. Matching the $1/e$ times is one choice of normalisation; any other choice that equates the two processes at a finite time leaves the same ordering for $t$ well below that time, because a quadratic onset is always flatter than a linear one there. A rate extracted from long-time telegraph noise therefore overstates the damage at short storage times when the underlying mechanism is a slowly fluctuating splitting, and the two descriptions can be distinguished experimentally by measuring the decay at $t\ll T^\star$ rather than fitting a single lifetime. This is the second half of Hypothesis~\ref{hyp:aniso}, and it has a direct consequence for hardware characterisation: the quantity to report is the transverse decay curve, not a single number.

Panel (c) converts the quasi-static channel into a requirement. With Tier~II parameters the asymptotic key rate vanishes for $\sigma_\delta$ above \SI{5.7e3}{rad/s} at \SI{50}{km} and above \SI{2.9e3}{rad/s} at \SI{100}{km}. In energy units these thresholds are $\hbar\sigma_\delta=\SI{3.7}{peV}$ and \SI{1.9}{peV}, so the root-mean-square residual hybridisation of the interrogated loop must stay in the picoelectronvolt range over the storage time. For comparison, reading the reported $\tau_X=\SI{14.5}{\micro s}$ as a coherent scale gives $\hbar/\tau_X=\SI{45}{peV}$, an order of magnitude above the requirement. Splittings are expected to fall exponentially with wire length \cite{albrecht2016exponential}, so this is a design target rather than a contradiction, but it is a target that is not usually stated in cryptographic terms.
\begin{figure*}[!t]
\centering
\includegraphics[width=0.92\textwidth]{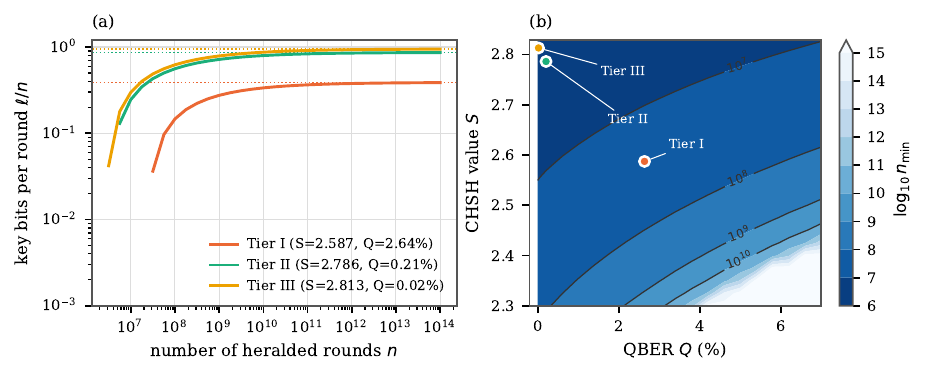}
\caption{(a) Finite-size key per kept round from Eq.~\eqref{eq:ell} with $\eps^{\mathrm s}=10^{-8}$ and optimised test fraction, for the \SI{50}{km} Barrett-Kok operating points; dotted lines show the asymptotic rates. (b) Minimum number of rounds with a positive key over the $(Q,S)$ plane, with the operating points marked.}
\label{fig:fs}
\end{figure*}

\begin{figure*}[!t]
\centering
\includegraphics[width=0.84\textwidth]{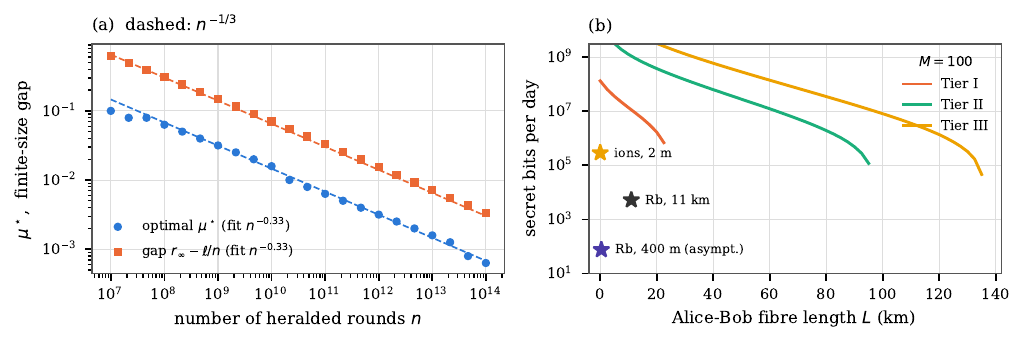}
\caption{(a) Optimal test fraction and finite-size penalty against the number of kept rounds at the Tier~II operating point, with fitted power laws and the $n^{-1/3}$ prediction of Proposition~\ref{prop:scaling}. (b) Secret bits per day against distance for $M=100$ memory pairs ($\eps^{\mathrm s}=10^{-8}$, first-order EAT), with the reported DI-QKD experiments shown as stars at their actual distances. The experiments use different security parameters and bounds (Rényi EAT with $\eps^{\mathrm s}=10^{-5}$ at \SI{11}{km}; asymptotic rate at \SI{400}{m}), so the comparison is indicative.}
\label{fig:sc}
\end{figure*}

\subsection{Finite-Size Requirements}
Fig.~\ref{fig:fs}a shows the key per round against the number of kept rounds for the \SI{50}{km} Barrett-Kok operating points, and Fig.~\ref{fig:fs}b maps the minimum round number over the $(Q,S)$ plane. Tier~I, at $S=2.587$ and $Q=\SI{2.64}{\percent}$, needs $n_{\min}=2.3\times10^{7}$; Tiers~II and III, at $S=2.786$ and $S=2.813$, need $3.2\times10^{6}$ and $2.7\times10^{6}$. The herald rates at the same distance are 0.36, 6.4 and \SI{45}{s^{-1}} per memory pair, so with one pair the accumulation times are $6.5\times10^{7}$, $5.0\times10^{5}$ and $6.0\times10^{4}$ seconds, that is about two years, six days and seventeen hours. In every case the time to a first key is set by the round requirement rather than by the asymptotic rate, which is the content of Hypothesis~\ref{hyp:finite}. Fig.~\ref{fig:sc}a verifies the scaling law of Proposition~\ref{prop:scaling}: fits over $n\ge10^9$ give $\mu^\star\propto n^{-0.334}$ and $r_\infty-\ell/n\propto n^{-0.330}$, both consistent with the predicted exponent $-1/3$. The practical consequence is that halving the finite-size penalty costs a factor of eight in rounds, so multiplexing and interface efficiency are more effective than further improvements in $S$ once $S$ exceeds about 2.7.
\begin{figure*}[!t]
\centering
\includegraphics[width=0.84\textwidth]{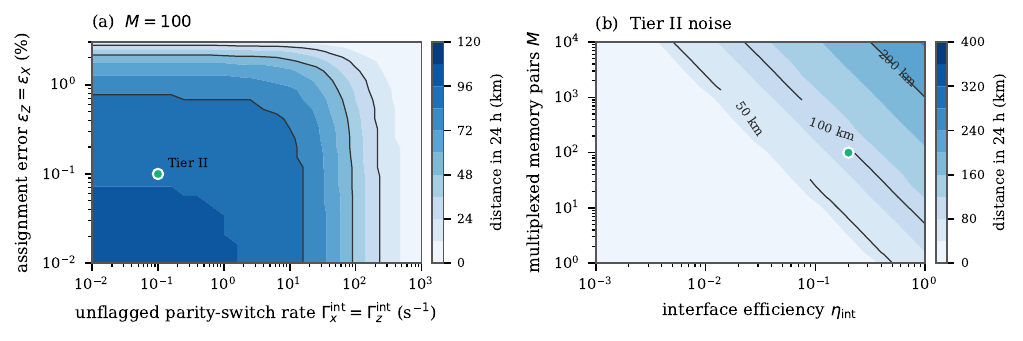}
\caption{Largest distance at which a finite-size key with $\eps^{\mathrm s}=10^{-8}$ is obtained within \SI{24}{h} using Barrett-Kok heralding. (a) Against the unflagged isotropic parity-switch rate and the assignment error, at Tier~II interface parameters and $M=100$. (b) Against interface efficiency and multiplexing, at Tier~II noise parameters. Markers show Tier~II.}
\label{fig:map}
\end{figure*}

\subsection{Operational Feasibility and Throughput}
Define $L_{24\mathrm h}$ as the largest distance at which $n_{\min}$ kept rounds are collected within \SI{24}{h}. Fig.~\ref{fig:map}a shows $L_{24\mathrm h}$ over the plane of unflagged switching rate and assignment error at Tier~II interface parameters with $M=100$ memory pairs. The reach saturates near \SI{96}{km} when the noise is low, remains at \SI{88}{km} for $\Gamma^{\mathrm{int}}_{x,z}=\SI{10}{s^{-1}}$ and at \SI{78}{km} for $\eps=\SI{1}{\percent}$, and falls to \SI{44}{km} at $\Gamma^{\mathrm{int}}_{x,z}=\SI{100}{s^{-1}}$ and to \SI{54}{km} at $\eps=\SI{2}{\percent}$. Fig.~\ref{fig:map}b shows that once the noise is below those levels the reach is set by the interface efficiency and the multiplexing: at $\etaint=0.2$ the values are \SI{24}{km}, \SI{96}{km} and \SI{138}{km} for $M=1$, $100$ and $1000$. Across tiers at $M=100$ the reach is \SI{24}{km}, \SI{96}{km} and \SI{136}{km}.
\begin{figure*}[!t]
\centering
\includegraphics[width=0.84\textwidth]{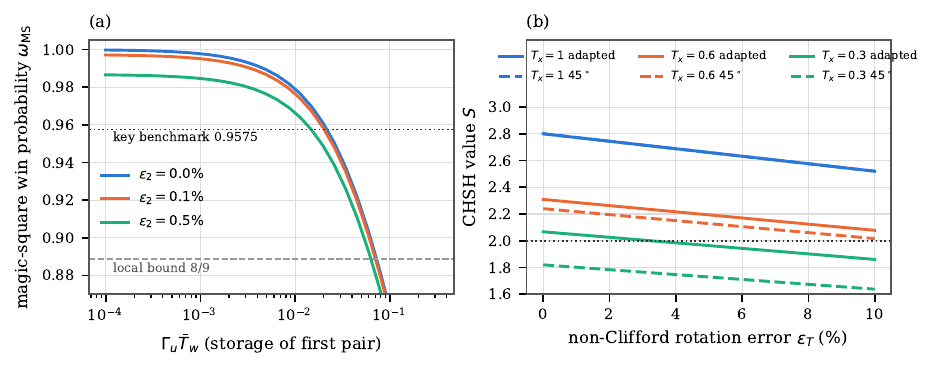}
\caption{(a) Magic-square winning probability against the storage-induced decay of the first pair for three assignment errors on the four-Majorana parities; the dotted line is the white-noise key benchmark of \cite{zhen2023device} and the dashed line the local bound $8/9$. (b) CHSH value against the non-Clifford rotation error for $a=0.99$ and three transverse correlations, with adapted (solid) and symmetric (dashed) settings.}
\label{fig:clifford}
\end{figure*}

Fig.~\ref{fig:sc}b expresses the same information as secret bits per day and places the published experiments on the same axes. At \SI{11}{km} with $M=100$, Tier~II gives $1.3\times10^{9}$ bits per day and Tier~III $1.0\times10^{10}$, against $5.2\times10^{3}$ bits per day extracted in the single-atom experiment at the same distance \cite{lu2026device}, $2.9\times10^{5}$ bits per day in the trapped-ion experiment at \SI{2}{m} \cite{nadlinger2022experimental} and $75$ bits per day, asymptotic, in the \SI{400}{m} atomic experiment \cite{zhang2022device}. These projections assume an interface efficiency and a degree of multiplexing that have not been demonstrated for Majorana devices, and the comparison should be read as a statement of what the hardware would have to deliver in order to be competitive rather than as a prediction of performance.

\subsection{Clifford-Native against Non-Clifford Nonlocality}
\label{sec:res_ms}
Fig.~\ref{fig:clifford}a shows the magic-square winning probability against the storage parameter $x=\Gamma_u\bar T_w$, where $\Gamma_u$ is the unflagged switching rate experienced by the stored pair. Reaching the indicative benchmark of \cite{zhen2023device} requires $x\lesssim0.024$ without readout error and $x\lesssim0.015$ at $\eps_2=\SI{0.5}{\percent}$; the necessary condition $\omega_{\mathrm{MS}}>8/9$ is met up to larger values of $x$, but it guarantees nonlocality, not a key. With no memory decay, assignment errors of \SI{0.1}{\percent} and \SI{0.5}{\percent} on the four-Majorana parities give winning probabilities of 0.9973 and 0.9868, and per-pair visibilities of 0.98 and 0.97 give 0.9846 and 0.9769. At the Tier~II operating point at \SI{50}{km} the average waiting time is \SI{0.16}{s} and $x=0.016$, which gives $\omega_{\mathrm{MS}}=0.966$, just above the benchmark; Tier~III gives $x=2.2\times10^{-5}$ and $\omega_{\mathrm{MS}}=0.997$. The magic square therefore trades the non-Clifford gate for the memory requirement $\Gamma_u\lesssim10^{-2}R_{\mathrm{att}}P_{\mathrm{kept}}$, and it produces at most half a key bit per entangled pair.

Fig.~\ref{fig:clifford}b quantifies the competing route. With a transverse correlation close to unity the CHSH value tolerates non-Clifford errors of several per cent, and under anisotropic noise the adapted settings of Proposition~\ref{prop:adapted} recover up to $0.25$ in $S$ relative to the symmetric choice. Taken together, the CHSH route is preferable whenever $\eps_T\lesssim\SI{1}{\percent}$ can be reached, while the magic square becomes attractive only for memories with $\Gamma_u\bar T_w\ll10^{-2}$, that is for devices whose unflagged switching rate is far below the herald rate.

\section{Discussion}
\label{sec:discussion}
\begin{table}[!tb]
\centering
\caption{Outcome of the five hypotheses of Sec.~\ref{sec:introduction}.}
\label{tab:hyp}
\begin{tabular}{@{}p{0.6cm}p{2.8cm}p{3.3cm}@{}}
\toprule
& Status & Evidence\\
\midrule
H\ref{hyp:clifford} & Established & Proposition~\ref{prop:nogo}; circumvented by Eq.~\eqref{eq:bob} or by the magic square\\
H\ref{hyp:flag} & Established in theory; flag not yet demonstrated & Lemmas~\ref{lem:flagdamp} and \ref{lem:flag}; \SI{5}{km} to 48 or \SI{233}{km} at $\Gqp=\SI{1000}{s^{-1}}$\\
H\ref{hyp:aniso} & Supported & Proposition~\ref{prop:adapted}, Corollary~\ref{cor:aniso}, Fig.~\ref{fig:qs}\\
H\ref{hyp:finite} & Supported & $n_{\min}$ and accumulation times, Proposition~\ref{prop:scaling}\\
H\ref{hyp:interface} & Supported within the model; needs an idle-memory measurement of $\tau_X$ & Tier~0 gives $S=1.93$, or a key below \SI{1}{km} under the most favourable reading; poisoning alone permits a key\\
\bottomrule
\end{tabular}
\end{table}

\subsection{Status of the Hypotheses}
Table~\ref{tab:hyp} summarises what the analysis establishes. Proof settles the first, and the second is established as a theorem whose premise, a fast total-parity measurement on a floating island, has yet to be demonstrated. Computation within the stated model supports the remaining three; the last will be decided by an idle-memory measurement of the transverse lifetime.
\begin{table}[!tb]
\centering
\caption{Reported DI-QKD demonstrations and the projections of this work.}
\label{tab:compare}
\setlength{\tabcolsep}{3pt}\begin{tabular}{@{}p{1.75cm}p{1.15cm}p{1.45cm}p{2.6cm}@{}}
\toprule
Platform & Distance & Security & Figure of merit\\
\midrule
Trapped ions \cite{nadlinger2022experimental} & \SI{2}{m} & finite-size & 95\,628 key bits from $1.5\times10^6$ pairs in \SI{8}{h}\\
Rb atoms \cite{zhang2022device} & \SI{400}{m} & asymptotic & $S=2.578$, $Q=\SI{7.8}{\percent}$, 0.07 bit per event\\
Photons \cite{liu2022toward} & $\le\SI{220}{m}$ & asymptotic & $\eta_{\mathrm{det}}=\SI{87.5}{\percent}$, positive rate\\
Rb atoms \cite{lu2026device} & 11 and \SI{100}{km} & finite-size and asymptotic & 0.112 bit per event at \SI{11}{km}, Rényi EAT\\
Majorana, Tier II (this work) & \SI{96}{km} & finite-size & key within \SI{24}{h} at $M=100$\\
\bottomrule
\end{tabular}
\end{table}

\subsection{Comparison with Other Platforms}
Table~\ref{tab:compare} places the projections next to the published demonstrations. Atomic and ionic systems have produced finite-size DI keys, and the single-atom experiment at \SI{100}{km} \cite{lu2026device} defines the current state of the art in reach. A Majorana node would have to match their fidelities, meaning $S\gtrsim2.6$ with a QBER of a few per cent, and in addition provide an optical interface that does not yet exist. Its potential advantages are a fast local parity readout, with a signal-to-noise ratio of one reached in a few microseconds \cite{microsoft2025interferometric} against state detection times of tens to hundreds of microseconds in trapped-ion and neutral-atom systems, and the prospect of dense on-chip multiplexing, which Fig.~\ref{fig:map}b identifies as the decisive lever. Neither advantage is automatic. Fast readout only helps while $M/(L/\cfib)$ exceeds the herald requirement, and dense multiplexing requires as many transduction channels as memories.

\subsection{Limitations}
The model assumes Markovian Poisson statistics for switching and poisoning, independent noise at the two nodes, and stochastic rather than coherent control errors; the quasi-static channel of Sec.~\ref{sec:memory} relaxes the first assumption only for the transverse component. A systematic over-rotation by $\delta$ in Eq.~\eqref{eq:bob} reduces $S^\star$ by $\cos\delta$ and is visible in the observed statistics, so it cannot compromise security, but it does consume noise budget. Correlated fluctuations of the parity, drive-induced heating that would couple the attempt rate to $\Gqp$, and cross-talk between multiplexed memories are outside the present treatment and would require device-specific modelling. The multiplexing factor $M$ assumes $M$ independent interfaces and readout chains; its cryogenic heat load and wiring cost grow linearly in $M$ and set a practical ceiling that our maps do not include. The flag assumes that the total island parity can be measured much faster than $1/\Gqp$ without back-action on the logical operators; the commutation relation guarantees the second property in the ideal case, and the first is a design constraint. Flag errors are covered by Eq.~\eqref{eq:phiimp}: false alarms cost only rate, while missed parity changes reintroduce the first-order loss in proportion to $\eps_m$, so the miss probability of the total-parity readout is a specification of the same importance as its speed. The interface parameters are hypothetical throughout, which is why the results are presented as requirement maps. Finally, the finite-size bound is conservative: the Rényi and generalised entropy accumulation theorems \cite{metger2024generalised}, improved entropy bounds \cite{brown2024device,tan2022improved} and advantage distillation \cite{tan2020advantage} would each reduce $n_{\min}$ and extend every distance quoted here.

\subsection{Falsifiable Predictions}
The analysis makes four predictions that can be tested before a full network node exists. A tetron whose transverse decay is measured at storage times well below the fitted lifetime will show the quadratic onset of Fig.~\ref{fig:qs}a if the dominant mechanism is a quasi-static splitting, and the linear onset if it is telegraph switching. A total-parity measurement interleaved with pair-parity readout will flag a fraction $\tfrac12(1-e^{-2\Gqp t})$ of rounds, and the coherence of the unflagged rounds will decay as $\cosh[(1-2u)\Gqp t]/\cosh(\Gqp t)$ rather than $e^{-2u\Gqp t}$, which measures the mode weights $u$ of Lemma~\ref{lem:flagdamp}; a residual decay that is linear in $t$ at short times measures the miss probability $\eps_m$ through Eq.~\eqref{eq:phiimpexp}. A two-node Bell test using braiding alone will not exceed $S=2$, whatever the fidelity of the braids. A CHSH test using one injected magic state per test round will, at the fidelities of Tier~I, give $S$ between 2.59 at \SI{50}{km} and 2.62 at zero distance, which is within reach of a laboratory-scale link once an interface exists.

\subsection{Outlook}
Because the flag decouples poisoning from distance, Majorana memories are natural candidates for the storage layer of memory-assisted architectures, where entanglement is generated over short segments and stored while other segments are attempted. The waiting-time trade-off quantified for the magic square in Fig.~\ref{fig:clifford}a reappears there as the central design parameter, with the unflagged switching rate playing the role of the memory figure of merit. Quantitative progress would follow from two developments: a demonstrated parity-to-photon interface with an efficiency of a few per cent, which would move Fig.~\ref{fig:map}b into the useful region, and the adoption of Rényi entropy accumulation, which would reduce the round requirements of Fig.~\ref{fig:fs} by orders of magnitude and shorten every accumulation time quoted in Sec.~\ref{sec:results}.

\section{Conclusion}
\label{sec:conclusion}
We have developed a framework for device-independent quantum key distribution with Majorana parity memories that begins with the Majorana operator algebra and measured parity-level quantities and ends with a composable finite-size key length. Four results shape the design of such a node. Clifford-only operations cannot certify CHSH nonlocality on heralded pairs, so a non-Clifford resource is required, and it can be confined to the test rounds of one party; the Pauli-native magic-square alternative replaces that resource with a demanding memory requirement. Quasiparticle poisoning changes the total island parity, and a just-in-time parity flag that is sound within the device-independent framework converts the poisoning-induced distance cutoff into a rate penalty and a second-order visibility loss, extending a near-term parameter set from \SI{5}{km} to between \SI{48}{km} and \SI{233}{km}, provided the flag misses fewer than about one per cent of parity changes in the most favourable case. Majorana noise is strongly anisotropic, adapted settings exploit that anisotropy, and the functional form of the transverse decay distinguishes quasi-static splitting from telegraph switching, with the quasi-static case setting a requirement at the picoelectronvolt level. Finally, the number of rounds demanded by the finite-size analysis, rather than the asymptotic rate, sets the time to a first key, and both the optimal test fraction and the finite-size penalty scale as $n^{-1/3}$.

Parameters inferred from 2025 tetron prototypes give $S=1.93$, below the local bound, under a Markovian reading of the data, and at most a sub-kilometre key under the most favourable reading, whereas hypothetical target parameters give finite-size keys over about \SI{100}{km} within a day when combined with an efficient photonic interface and of order $10^2$ multiplexed memories. The experimental milestones that follow are an $X$-basis parity memory time in the millisecond range, an $X$ assignment error near one per cent, a residual splitting of a few picoelectronvolts, a demonstrated total-parity flag, and a parity-to-photon interface. Until these are met, Majorana DI-QKD is best understood as a well-specified engineering target, and the framework presented here states that target in quantitative terms.

\bibliography{references}

\end{document}